\def\OPTIONConf{1}%
\newif\iftr\trtrue
\RequirePackage{etex}
\documentclass[9pt]{sigplanconf}
\toappear{}
\usepackage{amsmath}
\usepackage{amssymb}
\usepackage{amsthm}
\usepackage{ stmaryrd }
\usepackage{mathpartir}
\usepackage{balance}
\usepackage{stmaryrd}
\usepackage{wasysym}
\usepackage{multicol}
\usepackage{extarrows}

\usepackage[usenames,dvipsnames,svgnames,table]{xcolor}
\definecolor{light-gray}{gray}{0.9}
\usepackage{soul}
\setulcolor{red}
\usepackage{mathpazo}
\usepackage{colortab}
\usepackage[lowtilde]{url}
\usepackage{todonotes}
\usepackage{listings}
\usepackage{microtype}
\def \TirNameStyle #1{\small\rulename{#1}}
\renewcommand{\MathparLineskip}{\lineskiplimit=.3\baselineskip\lineskip=.3\baselineskip plus .2\baselineskip}

\usepackage{joshuadunfield}
\iftr

\else
	\usepackage{times}
\fi
\usepackage{llproof}
\usepackage{rulelinks}
\newtheorem{theorem}{Theorem}
\newtheorem{lemma}[theorem]{Lemma}

\newenvironment{proof-sketch}{\noindent{\emph{Proof Sketch.}}}{\qed}
\makeatletter

\usepackage[firstpage]{draftwatermark}
\SetWatermarkText{\hspace*{8in}\raisebox{6.5in}{\includegraphics[scale=0.1]{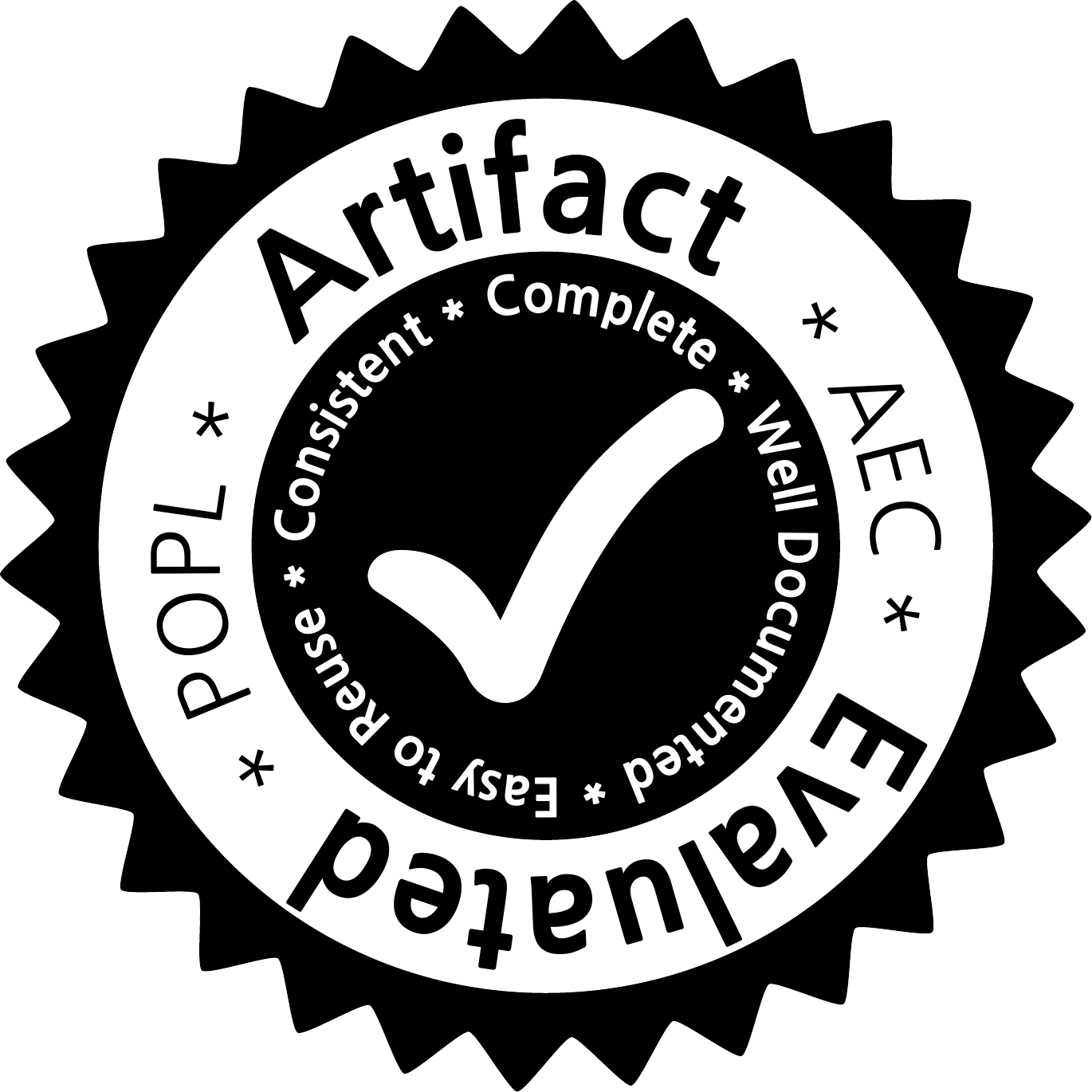}}}
\SetWatermarkAngle{0}

\renewcommand\topfraction{0.85}

\renewcommand\floatpagefraction{0.85}

\AtBeginDocument{%
 \abovedisplayskip=2pt plus 4pt
 \abovedisplayshortskip=0pt
 \belowdisplayskip=2pt plus 4pt
 \belowdisplayshortskip=0pt
}

\usepackage[compact]{titlesec}
\titlespacing*{\section}{0pt plus 3pt}{4pt plus 3pt}{2pt plus 3pt}
\titlespacing*{\subsection}{0pt plus 3pt}{4pt plus 3pt}{2pt plus 3pt}
\titlespacing*{\subsubsection}{0pt plus 3pt}{4pt plus 3pt}{2pt plus 3pt}
\titlespacing*{\paragraph}{0pt}{4pt}{3pt}
\usepackage[hidelinks,breaklinks,draft=false]{hyperref}

\usepackage{enumitem}
\makeatletter
\def\thm@space@setup{%
  \thm@preskip=4px plus 2px minus 2px
  \thm@postskip=\thm@preskip 
}
\makeatother

\usepackage{todonotes}
\usepackage{xcolor}

\newcommand{\llparenthesiscolor}{\textcolor{violet}{\llparenthesis}}
\newcommand{\rrparenthesiscolor}{\textcolor{violet}{\rrparenthesis}}

\newcommand{\htau}{\dot{\tau}}
\newcommand{\tarr}[2]{\inparens{#1 \rightarrow #2}}
\newcommand{\tarrnp}[2]{#1 \rightarrow #2}
\newcommand{\tnum}{\mathtt{num}}
\newcommand{\tehole}{\llparenthesiscolor\rrparenthesiscolor}
\newcommand{\tsum}[2]{\inparens{{#1} + {#2}}}

\newcommand{\tcompat}[2]{#1 \sim #2}
\newcommand{\tincompat}[2]{#1 \nsim #2}

\newcommand{\hexp}{\dot{e}}
\newcommand{\hlam}[2]{\inparens{\lambda #1.#2}}
\newcommand{\hap}[2]{#1(#2)}
\newcommand{\hnum}[1]{\underline{#1}}
\newcommand{\hadd}[2]{\inparens{#1 + #2}}
\newcommand{\hehole}{\llparenthesiscolor\rrparenthesiscolor}
\newcommand{\hhole}[1]{\llparenthesiscolor#1\rrparenthesiscolor}

\newcommand{\hinj}[2]{\mathtt{inj}_{#1}({#2})}
\newcommand{\hcase}[5]{\mathtt{case}({#1},{#2}.{#3},{#4}.{#5})}

\newcommand{\hGamma}{\dot{\Gamma}}
\newcommand{\domof}[1]{\text{dom}(#1)}
\newcommand{\hsyn}[3]{#1 \vdash #2 \Rightarrow #3}
\newcommand{\hana}[3]{#1 \vdash #2 \Leftarrow #3}

\newcommand{\zwsel}[1]{
  \setlength{\fboxsep}{0pt}
  \colorbox{green!10!white!100}{
    \ensuremath{{{\textcolor{Green}{{\hspace{-2px}\triangleright}}}}{#1}{\textcolor{Green}{\triangleleft{\vphantom{\tehole}}}}}}
}

\newcommand{\removeSel}[1]{#1^{\diamond}}

\newcommand{\ztau}{\hat{\tau}}

\newcommand{\zexp}{\hat{e}}

\newcommand{\dParent}{\mathtt{parent}}
\newcommand{\dChildn}[1]{\mathtt{child}~\mathtt{{#1}}}
\newcommand{\dChildnm}[1]{\mathtt{child}~{#1}}

\newcommand{\aMove}[1]{\mathtt{move}~#1}

\newcommand{\aDel}{\mathtt{del}}

\newcommand{\aConstruct}[1]{\mathtt{construct}~#1}
\newcommand{\aConstructx}[1]{#1}
\newcommand{\aFinish}{\mathtt{finish}}

\newcommand{\performAna}[5]{#1 \vdash #2 \xlongrightarrow{#4} #5 \Leftarrow #3}
\newcommand{\performAnaI}[5]{#1 \vdash #2 \xlongrightarrow{#4}\hspace{-3px}{}^{*}~ #5 \Leftarrow #3}
\newcommand{\performSyn}[6]{#1 \vdash #2 \Rightarrow #3 \xlongrightarrow{#4} #5 \Rightarrow #6}
\newcommand{\performSynI}[6]{#1 \vdash #2 \Rightarrow #3 \xlongrightarrow{#4}\hspace{-3px}{}^{*}~ #5 \Rightarrow #6}
\newcommand{\performTyp}[3]{#1 \xlongrightarrow{#2} #3}
\newcommand{\performTypI}[3]{#1 \xlongrightarrow{#2}\hspace{-3px}{}^{*}~#3}

\newcommand{\performMove}[3]{#1 \xlongrightarrow{#2} #3}

\newcommand{\farr}{\mathtt{arrow}}
\newcommand{\fnum}{\mathtt{num}}
\newcommand{\fsum}{\mathtt{sum}}

\newcommand{\fasc}{\mathtt{asc}}
\newcommand{\fvar}[1]{\mathtt{var}~#1}
\newcommand{\flam}[1]{\mathtt{lam}~#1}
\newcommand{\fap}{\mathtt{ap}}
\newcommand{\fnumlit}[1]{\mathtt{lit}~#1}
\newcommand{\fplus}{\mathtt{plus}}

\newcommand{\fnehole}{\mathtt{nehole}}

\newcommand{\finj}[1]{\mathtt{inj}~#1}
\newcommand{\fcase}[2]{\mathtt{case}~#1~#2}

\newcommand{\arrmatch}[2]{#1 \blacktriangleright_{\rightarrow} #2}

\newcommand{\TABperformAna}[5]{#1 \vdash & #2                & \xlongrightarrow{#4} & #5 & \Leftarrow #3}

\newcommand{\TABperformTyp}[3]{& #1 & \xlongrightarrow{#2} & #3}

\newcommand{\TABperformMove}[3]{#1 & \xlongrightarrow{#2} & #3}

\newcommand{\sumhasmatched}[2]{#1 \mathrel{\textcolor{black}{\blacktriangleright_{+}}} #2}

\newcommand{\subminsyn}[1]{\mathsf{submin}_{\Rightarrow}(#1)}
\newcommand{\subminana}[1]{\mathsf{submin}_{\Leftarrow}(#1)}

\newcommand{\inparens}[1]{{\color{gray}(}#1{\color{gray})}}

\newcommand{\rname}[1]{\textsc{#1}}
\newcommand{\gap}{\vspace{7pt}}

\begin{document}

\conferenceinfo{-}{-}
\copyrightyear{-}
\copyrightdata{[to be supplied]}

\preprintfooter{Draft}   

\title{Hazelnut: A Bidirectionally Typed \\ Structure Editor
 Calculus}
\iftr \subtitle{\vspace{-6px}Extended Version} \fi
\authorinfo{
        Cyrus Omar$^1$
        \and Ian Voysey$^1$
        \and Michael Hilton$^2$
        \and Jonathan Aldrich$^1$
        \and Matthew A. Hammer$^3$
}
{
 \begin{tabular}{ccc}
 	\begin{tabular}{c}
 $^1$Carnegie Mellon University, USA\\
 \textsf{\hphantom{$^1$}\{comar, iev, aldrich\}@cs.cmu.edu}
 \end{tabular} &
 \begin{tabular}{c}
 $^2$Oregon State University, USA\\
 \textsf{\hphantom{$^1$}hiltonm@eecs.oregonstate.edu}
 \end{tabular} &
  \begin{tabular}{c}
 $^3$University of Colorado Boulder, USA\\
 \textsf{\hphantom{$^1$}matthew.hammer@colorado.edu}
 \end{tabular}
 \end{tabular}
}
{
 \vspace{-12px}
}

\maketitle
\begin{abstract}

\emph{Structure editors} allow programmers to edit the tree structure of
a program directly. This can have cognitive benefits, particularly for
novice and end-user programmers. It also simplifies matters for tool
designers, because they do not need to contend with malformed program text.

This paper introduces Hazelnut, a {structure editor} based on a small
bidirectionally typed lambda calculus extended with \emph{holes} and
a \emph{cursor}. Hazelnut goes one step beyond
syntactic well-formedness: its {edit actions} operate over statically
meaningful incomplete terms.  Na\"ively, this would force the programmer to construct terms in a
rigid ``outside-in'' manner. To avoid this problem, the {action semantics}
automatically places terms assigned a type that is inconsistent with the
expected type \emph{inside} a {hole}. This meaningfully defers the type consistency
check until the term inside the hole is \emph{finished}.

Hazelnut is not intended as an end-user tool itself. Instead, it serves as a foundational account of typed structure
editing. To that end, we describe how
Hazelnut's rich metatheory, which we have mechanized using the Agda proof assistant, serves as a guide when we extend the calculus to include binary sum types. We also discuss various interpretations of holes, and in so doing
reveal connections with gradual typing and contextual modal type theory, the Curry-Howard interpretation of contextual modal logic. Finally, we
discuss how Hazelnut's semantics lends itself to implementation as an event-based functional reactive program. Our simple reference implementation is written using \lstinline{js_of_ocaml}.

\end{abstract}


\section{Introduction}\label{sec:introduction}

Programmers typically interact with meaningful programs only
indirectly, by editing text that is first parsed according to a textual
syntax and then typechecked according to a static semantics. This
indirection has practical benefits, to be sure -- text editors and other
text-based tools benefit from decades of development effort. However, this indirection through text also
introduces some fundamental complexity into the programming process.

First, it requires that programmers learn the subtleties of the textual
syntax (e.g. operator precedence.) This is particularly 
challenging for novices \cite{Altadmri:2015:MCI:2676723.2677258,stefik2013empirical,Marceau:2011:VGT:2016911.2016921}, and even experienced programmers frequently make mistakes \cite{7548903,jones2006developer,Marceau:2011:VGT:2016911.2016921}.

Second, many sequences of characters do not correspond to meaningful
programs. This complicates the design of program editors and other interactive programming tools. 
In a dataset  
gathered by Yoon and Myers consisting of 1460 hours of Eclipse edit logs  \cite{6883030}, 44.2\% of
edit states were syntactically malformed. Some additional
percentage of edit states were well-formed but ill-typed (the dataset was
not complete enough to determine the exact percentage.) Collectively, we refer to these edit states as \emph{meaningless edit states}, because they are not given static or dynamic meaning by the language definition. 
As a consequence, it is difficult
to design useful language-aware editor services, e.g. syntax
highlighting~\cite{sarkar2015impact}, type-aware code
completion~\cite{Mooty:2010:CCC:1915084.1916348,Omar:2012:ACC:2337223.2337324},
and refactoring services \cite{mens2004survey}. Editors must either disable
these editor services when they encounter meaningless edit states or
deploy \emph{ad hoc} heuristics, e.g. by using whitespace to guess the intent \cite{DBLP:conf/oopsla/KatsJNV09,DBLP:conf/sle/JongeNKV09}.

These complications have motivated a long line of research
into \emph{structure editors}, i.e. program editors where every edit state
corresponds to a program structure \cite{teitelbaum_cornell_1981}.\footnote{Structure editors are also variously known as \emph{structured editors}, \emph{structural editors}, \emph{syntax-directed editors} and \emph{projectional editors}.}

Most structure editors are \emph{syntactic structure editors}, i.e. the
edit state corresponds to a syntax tree with \emph{holes} that stand for branches of
the tree that have yet to be constructed, and the edit actions are
context-free tree transformations. For example, Scratch is a syntactic
structure editor that has achieved success as a tool for teaching children
how to program \cite{Resnick:2009:SP:1592761.1592779}. 

Researchers have also 
 designed  syntactic structure editors for more sophisticated languages with non-trivial  
static type systems. Contemporary examples include \texttt{mbeddr}, a structure editor for a C-like
language \cite{voelter_mbeddr:_2012}, TouchDevelop, a structure editor for an
object-oriented language \cite{tillmann_touchdevelop:_2011}, and Lamdu, a structure 
editor for a functional language similar to Haskell \cite{lamdu}. Each of
these editors presents an innovative user interface, but the non-trivial
type and binding structure of the underlying language complicates its
design. The problem is that syntactic structure editors do not assign static meaning to every edit state -- they guarantee only that every edit state corresponds to  a 
syntactically well-formed tree. These editors must also either selectively disable editor services that require an understanding of the semantics of the program being written, or deploy \emph{ad hoc} heuristics.

This paper develops a principled solution to this problem. We introduce
Hazelnut, a \emph{typed structure editor} based on a bidirectionally typed
lambda calculus extended to assign static meaning to expressions and types
with {holes}, which we call \textbf{H-expressions}
and \textbf{H-types}. Hazelnut's formal \emph{action semantics} maintains
the invariant that every edit state is a statically meaningful
(i.e. well-H-typed) H-expression with a single superimposed \emph{cursor}. We
call H-expressions and H-types with a cursor \textbf{Z-expressions}
and \textbf{Z-types} (so prefixed because our encoding follows
Huet's \emph{zipper} pattern \cite{JFP::Huet1997}.)

Na\"ively, enforcing an injunction on ill-H-typed edit states would force
programmers to construct programs in a rigid ``outside-in'' manner. For
example, the programmer would often need to construct the outer function
application form before identifying the intended function. To address this
problem, Hazelnut leaves newly constructed expressions \emph{inside} a hole
if the expression's type is inconsistent with the expected type. This
meaningfully defers the type consistency check until the expression inside
the hole is \emph{finished}. In other words, holes appear both at the
leaves and at the internal nodes of an H-expression that remain under
construction.

The remainder of this paper is organized as follows:
\begin{itemize}[itemsep=0px,partopsep=2px,topsep=2px]
  \item We begin in Sec.  \ref{sec:example} with two examples of edit
    sequences to develop the reader's intuitions.

  \item We then give a detailed overview of Hazelnut's semantics and
  metatheory, which has been mechanized using the Agda proof assistant, in
  Sec.  \ref{sec:hazel}.

  \item Hazelnut is a {foundational} calculus, i.e. a calculus that
  language and editor designers are expected to extend with higher level
  constructs. We extend Hazelnut with binary sum types in
  Sec.  \ref{sec:extending} to demonstrate how Hazelnut's rich metatheory
  guides one such extension.

  \item In Sec.  \ref{sec:impl}, we briefly describe how Hazelnut's action
  semantics lends itself to efficient implementation as a functional
  reactive program. Our reference implementation is written
  using the OCaml \lstinline{React} library and \lstinline{js_of_ocaml}.

  \item In Sec.  \ref{sec:rw}, we summarize related work. In particular,
  much of the technical machinery needed to handle type holes coincides
  with machinery developed for gradual type systems. Similarly,
  expression holes can be interpreted as the closures of contextual modal
  type theory, which, by its correspondence with contextual modal logic,
  suggests logical foundations for the system.

  \item We conclude in Sec.  \ref{sec:future} by summarizing our vision of
  a principled science of structure editor design rooted in type theory,
  and suggest a number of future directions.
\end{itemize}
The supplemental material can be accessed from:
\[\text{\url{http://cs.cmu.edu/~comar/hazelnut-popl17/}} \]

\section{Programming in Hazelnut}\label{sec:example}
\subsection{Example 1: Constructing the Increment Function}

Figure~\ref{fig:first-example} shows an edit sequence that constructs the
increment function, of type $\tarr{\tnum}{\tnum}$, starting from the empty
hole via the indicated sequence of {actions}. We will introduce Hazelnut's
formal syntax and define the referenced rules in
Sec. \ref{sec:hazel}.\footnote{For concision, the column labeled \textbf{Rule} in Figures \ref{fig:first-example} and \ref{fig:second-example} 
indicates only the relevant \emph{non-zipper} rule (see Sec. \ref{sec:zipper-cases}.) The reader is encouraged to follow along with these examples using the reference implementation. The derivations for these examples are also available in the Agda mechanization.} First, let us build some high-level intuitions.

The edit state in Hazelnut is a {Z-expression}, $\zexp$. Every Z-expression
has a single {H-expression}, $\hexp$, or {H-type}, $\htau$, under the
{cursor}, typeset $\zwsel{\hexp}$ or $\zwsel{\htau}$, respectively.\footnote{The reference implementation omits the triangles, while the Agda mechanization necessarily omits the colors.} For
example, on Line 1, the empty expression hole, $\hhole{}$, is under the
cursor.

Actions act relative to the cursor. The first action, on Line 1, is
$\aConstruct{\flam{x}}$. This results in the Z-expression on
Line 2, consisting of a lambda abstraction with argument $x$ under an arrow type
ascription. The cursor is placed (arbitrarily) on the argument type hole.

The actions on Lines 2-5 complete the type ascription. In particular, the
$\aConstruct{\fnum}$ action constructs the $\tnum$ type at the cursor and
the $\aMove{\dParent}$ and $\aMove{\dChildn{2}}$ action sequence 
moves the cursor to the next hole. In practice, an editor would also define compound movement
actions, e.g. an action that moves the cursor directly to the next
 hole, in terms of these primitive movement actions.

\begin{figure}[t!]
  \label{ex1}
\begin{center}
$\arraycolsep=2px
\begin{array}[p]{|c||l|l||l|l|}
\hline
\# & \textbf{Z-Expression} &
\textbf{Next Action} & \textbf{Rule}
\\
\hline
1 &
\zwsel{\hhole{}} &
\aConstruct{\flam{x}} &
\text{(\ref{r:conelamhole})}
\\ 2 &
{\hlam{x}{\hhole{}}} : \tarr{\zwsel{\hhole{}}}{\hhole{}} &
\aConstruct{\fnum{}} &
\text{(\ref{r:contnum})}
\\ 3 &
{\hlam{x}{\hhole{}}} : \tarr{\zwsel{\tnum{}}}{\hhole{}} &
\aMove{\dParent} &
\text{({\ref{rule:move-parent-arr-left}})}
\\ 4 &
{\hlam{x}{\hhole{}}} : \zwsel{\tarr{\tnum{}}{\hhole{}}} &
\aMove{\dChildn{2}} &
\text{({\ref{rule:move-arr-c2}})}
\\ 5 &
{\hlam{x}{\hhole{}}} : \tarr{\tnum}{\zwsel{\hhole{}}}
&
\aConstruct{\fnum{}} &
\text{(\ref{r:contnum})}
\\ 6 &
{\hlam{x}{\hhole{}}} : \tarr{\tnum{}}{\zwsel{\tnum{}}} &
\aMove{\dParent{}} &
\text{(\ref{rule:move-parent-arr-right})}
\\ 7 &
{\hlam{x}{\hhole{}}} : \zwsel{\tarr{\tnum{}}{\tnum{}}}
&
\aMove{\dParent{}} &
\text{(\ref{rule:move-parent-asc-right})}
\\ 8 &
\zwsel{{\hlam{x}{\hhole{}}} : \tarr{\tnum}{\tnum}} &
\aMove{\dChildn{1}} &
\text{(\ref{r:movefirstchild-asc})}
\\ 9 &
{\zwsel{\hlam{x}{\hhole{}}}} : \tarr{\tnum{}}{\tnum{}} &
\aMove{\dChildn{1}} &
\text{(\ref{r:movefirstchild-lam})}
\\ 10 &
{\hlam{x}{\zwsel{\hhole{}}}} : \tarr{\tnum{}}{\tnum{}} &
\aConstruct{\fvar{x}} &
\text{(\ref{r:conevar})}
\\ 11 &
{\hlam{x}{\zwsel{{x}}}} : \tarr{\tnum{}}{\tnum{}} &
{\aConstruct{\fplus}}
&
\text{(\ref{rule:construct-plus-compat})}
\\ 12 &
{\hlam{x}{\hadd{x}{\zwsel{\hhole{}}}}} : \tarr{\tnum{}}{\tnum{}} &
\aConstruct{\fnumlit{1}} &
\text{(\ref{r:conenumnum})}
\\ 13 &
{\hlam{x}{\hadd{x}{\zwsel{\hnum{1}}}}} : \tarr{\tnum}{\tnum} &
\textrm{---} &
{\textrm{---}}
\\ \hline
\end{array}
$\end{center}\vspace{-6px}
\caption{Constructing the increment function in Hazelnut.}
\label{fig:first-example}
\end{figure}

After moving to the function body, Lines 10-12 complete it by first constructing the variable
$x$, then constructing the plus form, and finally constructing the number literal 
$\hnum{1}$. Notice that we did not need to construct the function body in
an ``outside-in'' manner, i.e. we constructed $x$ before constructing the
outer plus form inside which $x$ ultimately appears. The transient function bodies, $x$ and $\hadd{x}{\hhole{}}$, can be checked against the given return type, $\tnum$ (as we will detail in
Sec. \ref{sec:holes}.)

\subsection{Example 2: Applying the Increment Function}

\begin{figure}[t!]
  \label{ex2}
\begin{center}
\colorbox{light-gray}{\hspace{54px} now assume $incr : \tarrnp{\tnum}{\tnum}$ \hspace{53px}}
$\arraycolsep=4px
\begin{array}{|r||l|l||l|l|}
\hline
\# & \textbf{Z-Expression} &
\textbf{Next Action} & \textbf{Rule}
\\
\hline
14 &
\zwsel{\hhole{}} &
\aConstruct{\fvar{incr}} \hphantom{~\,\,~~}&
\text{(\ref{r:conevar})}
\\ 15 &
\zwsel{incr} &
\aConstruct{\fap} &
\text{(\ref{r:coneapfn})}
\\ 16 &
incr(\zwsel{\hhole{}}) &
\aConstruct{\fvar{incr}} &
\text{(\ref{r:conevar2})}
\\ 17 &
incr(\hhole{\zwsel{incr}}) &
\aConstruct{\fap} &
\text{(\ref{r:coneapfn})}
\\ 18 &
incr(\hhole{incr(\zwsel{\hhole{}})}) \hphantom{~~~~} &
\aConstruct{\fnumlit{3}} &
\text{(\ref{r:conenumnum})}
\\ 19 &
incr(\hhole{incr(\zwsel{\hnum{3}})}) &
\aMove{\dParent}&
\text{(\ref{r:moveparent-ap2})}
\\ 20 &
incr(\hhole{\zwsel{incr(\hnum{3})}}) &
\aMove{\dParent} &
\text{(\ref{r:moveparent-hole})}
\\ 21 &
incr(\zwsel{\hhole{incr(\hnum{3})}})&
\aFinish &
\text{(\ref{r:finishana})}
\\ 22 &
incr(\zwsel{incr(\hnum{3})}) &
\textrm{---} &
{\textrm{---}}
\\ \hline
\end{array}
$\end{center}\vspace{-6px}
\caption{Applying the increment function.}
\label{fig:second-example}
\end{figure}

Figure \ref{fig:second-example} shows an edit sequence that constructs the
expression $incr(incr(\hnum{3}))$, where $incr$ is assumed bound to the
increment function from Figure \ref{fig:first-example}.

We begin on Line 14 by constructing the variable $incr$. Line 15 then
constructs the application form with $incr$ in function position, leaving
the cursor on a hole in the argument position. Notice again that we did not construct the outer application form before identifying the function
being applied. 

We now need to apply $incr$ again, so we perform the same action on Line 16
as we did on Line 14, i.e. $\aConstruct{\fvar{incr}}$. In a syntactic
structure editor, performing such an action would result in the following
edit state:
\[
incr(\zwsel{incr})
\]
This edit state is ill-typed (after \emph{cursor erasure}): the argument of
$incr$ must be of type $\tnum$ but here it is of type
$\tarr{\tnum}{\tnum}$. Hazelnut does not allow such an edit state to arise.

We could alternatively have performed the $\aConstruct{\fap}$
action on Line 16. This would result in the following edit state, which is
well-typed according to the static semantics that we will define in the
next section:
\[
incr(\hhole{}(\zwsel{\hhole{}}))
\]
The problem is that the programmer is not able to identify the intended
function, $incr$, before constructing the function application form. This stands in
contrast to Lines 14-15.

Hazelnut's action semantics addresses this problem: rather than disallowing
the $\aConstruct{\fvar{incr}}$ action on Line 16, it leaves $incr$ inside a
hole:
\[
incr(\hhole{\zwsel{incr}})
\]
This defers the type consistency check, exactly as an empty hole in the
same position does. One way to think about non-empty holes is as an
internalization of the ``squiggly underline'' that text or syntactic
structure editors display to indicate a type inconsistency. By
internalizing this concept, the presence of a type inconsistency does not
leave the entire program formally meaningless.

The expression inside a non-empty hole must itself be well-typed, so the
programmer can continue to edit it. Lines 17-18 proceed to apply the inner
mention of $incr$ to a number literal, $\hnum{3}$. Finally, Lines 18-19
move the cursor to the non-empty hole and Line 21 performs the $\aFinish$
action. The $\aFinish$ action removes the hole if the type of the
expression inside the hole is consistent with the expected type, as it now
is. This results in the final edit state on Line 22, as desired. In
practice, the editor might automatically perform the $\aFinish$ action as
soon as it becomes possible, but for simplicity, Hazelnut formally requires
that it be performed explicitly.

\section{Hazelnut, Formally}
\label{sec:hazel}

The previous section introduced Hazelnut by example. In this section, we
systematically define the following structures:
\begin{itemize}[itemsep=0px,partopsep=2px,topsep=2px]
\item \textbf{H-types} and \textbf{H-expressions} (Sec. \ref{sec:holes}),
  which are types and expressions with {holes}. H-types classify
  H-expressions according to Hazelnut's \textbf{bidirectional static
    semantics}.

\item \textbf{Z-types} and \textbf{Z-expressions} (Sec. \ref{sec:cursors}),
  which superimpose\- a \emph{cursor} onto H-types and H-expressions,
  respectively (following Huet's \emph{zipper pattern}
  \cite{JFP::Huet1997}.) Every Z-type (resp. Z-expression) corresponds to
  an H-type (resp. H-expression) by \emph{cursor erasure}.

\item \textbf{Actions} (Sec. \ref{sec:actions}), which act relative to the
  cursor according to Hazelnut's \textbf{bidirectional action
    semantics}. The action semantics enjoys a rich metatheory. Of
  particular note, the \emph{sensibility theorem} establishes that every
  edit state is well-typed after cursor erasure.
\end{itemize}

Our overview below omits certain ``uninteresting'' details. The supplement
includes the complete collection of rules, in definitional order. These
rules, along with the proofs of all of the metatheorems discussed in this section (and several omitted auxiliary lemmas), have been
mechanized using the Agda proof assistant \cite{norell:thesis} (discussed in Sec. \ref{sec:mech}.)

\subsection{H-types and H-expressions}\label{sec:holes}
Figure \ref{fig:hexp-syntax} defines the syntax of H-types, $\htau$, and
H-expressions, $\hexp$. Most forms correspond directly to those of the
simply typed lambda calculus (STLC) extended with a single base type,
$\tnum$, of numbers (cf. \cite{pfpl}.) The number expression corresponding
to the mathematical number $n$ is drawn $\hnum{n}$, and for simplicity, we
define only a single arithmetic operation, $\hadd{\hexp}{\hexp}$. The form
$\hexp : \htau$ is an explicit \emph{type ascription}. In addition to these standard forms, \emph{type holes} and \emph{empty
  expression holes} are both drawn $\hehole$ and \emph{non-empty expression
  holes} are drawn $\hhole{\hexp}$.
Types and expressions that contain no holes are \emph{complete types} and
\emph{complete expressions}, respectively.

\begin{figure}[t]
$\arraycolsep=4pt\begin{array}{lllllll}
\mathsf{HTyp} & \htau & ::= &
  \tarr{\htau}{\htau} ~\vert~
  \tnum ~\vert~
  \tehole\\
\mathsf{HExp} & \hexp & ::= &
  x ~\vert~
  {\hlam{x}{\hexp}} ~\vert~
  \hap{\hexp}{\hexp} ~\vert~
  \hnum{n} ~\vert~
  \hadd{\hexp}{\hexp} ~\vert~
  \hexp : \htau ~\vert~
  \hehole ~\vert~
  \hhole{\hexp}
\end{array}$
\caption{Syntax of H-types and H-expressions. Metavariable $x$ ranges over variables and $n$ ranges over numerals.}
\label{fig:hexp-syntax}
\end{figure}

\begin{figure}
\fbox{$\hsyn{\hGamma}{\hexp}{\htau}$}~~\text{$\hexp$ synthesizes $\htau$}
\begin{subequations}
\begin{equation}\label{rule:syn-var}
\inferrule{ }{
  \hsyn{\hGamma, x : \htau}{x}{\htau}
}
\end{equation}
\begin{equation}\label{rule:syn-ap}
\inferrule{
  \hsyn{\hGamma}{\hexp_1}{\htau_1}\\
  \arrmatch{\htau_1}{\tarr{\htau_2}{\htau}}\\
  \hana{\hGamma}{\hexp_2}{\htau_2}
}{
  \hsyn{\hGamma}{\hap{\hexp_1}{\hexp_2}}{\htau}
}
\end{equation}
\begin{equation}\label{rule:syn-num}
\inferrule{ }{
  \hsyn{\hGamma}{\hnum{n}}{\tnum}
}
\end{equation}
\begin{equation}\label{rule:syn-plus}
\inferrule{
  \hana{\hGamma}{\hexp_1}{\tnum}\\
  \hana{\hGamma}{\hexp_2}{\tnum}
}{
  \hsyn{\hGamma}{\hadd{\hexp_1}{\hexp_2}}{\tnum}
}
\end{equation}
\begin{equation}\label{rule:syn-asc}
\inferrule{
  \hana{\hGamma}{\hexp}{\htau}
}{
  \hsyn{\hGamma}{\hexp : \htau}{\htau}
}
\end{equation}
\begin{equation}\label{rule:syn-ehole}
\inferrule{ }{
  \hsyn{\hGamma}{\hehole}{\tehole}
}
\end{equation}
\begin{equation}\label{rule:syn-hole}
\inferrule{
  \hsyn{\hGamma}{\hexp}{\htau}
}{
  \hsyn{\hGamma}{\hhole{\hexp}}{\tehole}
}
\end{equation}
\end{subequations}
\fbox{$\hana{\hGamma}{\hexp}{\htau}$}~~\text{$\hexp$ analyzes against $\htau$}
\begin{subequations}
\begin{equation}\label{rule:syn-lam}
\inferrule{
  \arrmatch{\htau}{\tarr{\htau_1}{\htau_2}}\\
  \hana{\hGamma, x : \htau_1}{\hexp}{\htau_2}
}{
  \hana{\hGamma}{\hlam{x}{\hexp}}{\htau}
}
\end{equation}
\begin{equation}\label{rule:ana-subsume}
\inferrule{
  \hsyn{\hGamma}{\hexp}{\htau'}\\
  \tcompat{\htau}{\htau'}
}{
  \hana{\hGamma}{\hexp}{\htau}
}
\end{equation}
\end{subequations}
\caption{H-type synthesis and analysis.}
\label{fig:ana-syn}
\end{figure}
\begin{figure}
\vspace{-1px}\noindent\fbox{$\tcompat{\htau}{\htau'}$}~~\text{$\htau$ and $\htau'$ are consistent}\vspace{-4px}
\begin{subequations}\label{eqns:consistency}
\begin{mathpar}
\inferrule{ }{
  \tcompat{\tehole}{\htau}
}
~~~~~
\inferrule{ }{
  \tcompat{\htau}{\tehole}
}
~~~~~
\inferrule{ }{
  \tcompat{\htau}{\htau}
}
~~~~~
\inferrule{
  \tcompat{\htau_1}{\htau_1'}\\
  \tcompat{\htau_2}{\htau_2'}
}{
  \tcompat{\tarr{\htau_1}{\htau_2}}{\tarr{\htau_1'}{\htau_2'}}
}~~~~\text{\hspace{-2px}(\ref*{eqns:consistency}a-d)}
\end{mathpar}
\end{subequations}
\noindent\fbox{$\arrmatch{\htau}{\tarr{\htau_1}{\htau_2}}$}~~\text{$\htau$ has matched arrow type $\tarr{\htau_1}{\htau_2}$}\vspace{-4px}

\begin{subequations}
\begin{minipage}{0.43\linewidth}
\begin{equation}
\inferrule{ }{
  \arrmatch{\tehole}{\tarr{\tehole}{\tehole}}
}
\end{equation}
\end{minipage}
\begin{minipage}{0.55\linewidth}
\begin{equation}
\inferrule{ }{
  \arrmatch{\tarr{\htau_1}{\htau_2}}{\tarr{\htau_1}{\htau_2}}
}
\end{equation}
\end{minipage}
\end{subequations}
\caption{H-type consistency and matched arrow types.}
\label{fig:type-consistency}
\end{figure}

Hazelnut's static semantics is organized as a \emph{bidirectional type
  system}
\cite{Pierce:2000:LTI:345099.345100,DBLP:conf/icfp/DaviesP00,DBLP:conf/tldi/ChlipalaPH05,bidi-tutorial}
around the two mutually defined judgements in Figure
\ref{fig:ana-syn}. Derivations of the type analysis judgement, $\hana{\hGamma}{\hexp}{\htau}$,
establish that $\hexp$ can appear where an expression of type $\htau$ is
expected. Derivations of the type synthesis judgement,
$\hsyn{\hGamma}{\hexp}{\htau}$, synthesize (a.k.a. \emph{locally infer} \cite{Pierce:2000:LTI:345099.345100}) a type from $\hexp$. Type synthesis is
necessary in positions where an expected type is not available (e.g. at the
top level.) Algorithmically, the type is an ``input'' of the type analysis
judgement, but an ``output'' of the type synthesis judgement.  Making a
judgemental distinction between these two notions will be essential in our
action semantics (Sec. \ref{sec:actions}.)

If an expression is
able to synthesize a type, it can also be analyzed against that type, or
any other \emph{consistent} type, according to the \emph{subsumption rule},
Rule (\ref{rule:ana-subsume}).

The \emph{H-type consistency judgement}, $\tcompat{\htau}{\htau'}$, that
appears as a premise in the subsumption rule is a reflexive and symmetric
(but not transitive) relation between H-types defined judgementally in
Figure \ref{fig:type-consistency}. This relation coincides with equality
for complete H-types. Two incomplete H-types are consistent if they differ
only at positions where a hole appears in either type. The type hole is
therefore consistent with every type. This notion of H-type consistency
coincides with the notion of type consistency that Siek and Taha discovered
in their foundational work on gradual type systems, if we interpret the
type hole as the $?$ (i.e. unknown) type \cite{Siek06a}.

Typing contexts, $\hGamma$, map each variable $x \in
\domof{\hGamma}$ to an hypothesis $x : \htau$. We
identify contexts up to exchange and contraction and adopt the standard identification
convention for structures that differ only up to alpha-renaming of bound variables. All hypothetical judgements obey a standard weakening lemma. Rule (\ref{rule:syn-var}) establishes that variable expressions synthesize the hypothesized H-type, in the standard manner.

Rule (\ref{rule:syn-lam}) defines analysis for lambda abstractions,
$\hlam{x}{\hexp}$. There is no type synthesis rule that applies to this
form, so lambda abstractions can appear only in analytic position,
i.e. where an expected type is known.\footnote{It is possible to also
  define a ``half-annotated'' synthetic lambda form, $\lambda x{:}\tau.e$,
  but for simplicity, we leave it out \cite{DBLP:conf/tldi/ChlipalaPH05}.}
Rule (\ref{rule:syn-lam}) is not quite the standard rule, as reproduced below:
\begin{equation*}
\inferrule{
  \hana{\hGamma, x : \htau_1}{\hexp}{\htau_2}
}{
  \hana{\hGamma}{\hlam{x}{\hexp}}{\tarr{\htau_1}{\htau_2}}
}
\end{equation*}
The problem is that this standard rule alone leaves us unable to analyze
lambda abstractions against the type hole, because the type hole is not
immediately of the form $\tarr{\htau_1}{\htau_2}$. There are two plausible
solutions to this problem. One solution would be to define a second rule
specifically for this case:
\begin{equation*}
\inferrule{
  \hana{\hGamma, x : \tehole}{\hexp}{\tehole}
}{
  \hana{\hGamma}{\hlam{x}{\hexp}}{\tehole}
}
\end{equation*}
Instead, we combine these two possible rules into a single rule through the
simple auxiliary \emph{matched arrow type} judgement,
$\arrmatch{\htau}{\tarr{\htau_1}{\htau_2}}$, defined in Figure
\ref{fig:type-consistency}. This judgement leaves arrow types alone and
assigns the type hole the matched arrow type $\tarr{\tehole}{\tehole}$. It
is easy to see that the two rules above are admissible by appeal to Rule
(\ref{rule:syn-lam}) and the matched arrow type judgement.  Encouragingly,
the matched arrow type judgement also arises in gradual type systems
\cite{DBLP:conf/popl/CiminiS16,DBLP:conf/popl/GarciaC15,DBLP:conf/popl/RastogiCH12}.

Rule (\ref{rule:syn-ap}) is again nearly the standard rule for function
application. It also makes use of the matched function type judgement to
combine what would otherwise need to be two rules for function application
-- one for when $e_1$ synthesizes an arrow type, and another for when $e_1$
synthesizes $\tehole$.

Rule (\ref{rule:syn-num}) states that numbers synthesize the $\tnum$
type. Rule (\ref{rule:syn-plus}) states that $\hexp_1 + \hexp_2$ behaves
like a function over numbers.

Rule (\ref{rule:syn-asc}) defines type synthesis of expressions of
ascription form, $\hexp : \htau$. This allows the programmer to explicitly state
a type for the ascribed expression to be analyzed against.

The rules described so far are sufficient to type complete
H-expressions. The two remaining rules give H-expressions with holes a
well-defined static semantics.

Rule (\ref{rule:syn-ehole}) states that the empty expression hole
synthesizes the type hole. Non-empty holes, which contain an H-expression that is ``under construction''
as described in Sec. \ref{sec:example}, also synthesize the hole type. According to Rule
(\ref{rule:syn-hole}), the enveloped expression must synthesize some (arbitrary) type. (We do not need non-empty type holes because every H-type is a valid classifier of H-expressions.)

Because the hole type is consistent with every type, expression holes can be analyzed against any type by subsumption. For example, it is instructive to derive the following:
\[\hana{incr : \tarr{\tnum}{\tnum}}{\hhole{incr}}{\tnum}\]

\subsection{Z-types and Z-expressions}\label{sec:cursors}
\newcommand{\cvert}{{\,{\vert}\,}}
\begin{figure}[t]
$\arraycolsep=2pt\begin{array}{lllllll}
\mathsf{ZTyp} & \ztau & ::= &
  \zwsel{\htau} \cvert
  \tarr{\ztau}{\htau} \cvert
  \tarr{\htau}{\ztau} \\
\mathsf{ZExp} & \zexp & ::= &
  \zwsel{\hexp} \cvert
  \hlam{x}{\zexp} \cvert
  \hap{\zexp}{\hexp} \cvert
  \hap{\hexp}{\zexp} \cvert
  \hadd{\zexp}{\hexp} \cvert
  \hadd{\hexp}{\zexp} \\
& & \cvert &
    \zexp : \htau \cvert
  \hexp : \ztau \cvert
  \hhole{\zexp}
\end{array}$
\caption{Syntax of Z-types and Z-expressions.}
\label{fig:zexp-syntax}
\end{figure}

Figure \ref{fig:zexp-syntax} defines the syntax of
Z-types, $\ztau$, and Z-expressions, $\zexp$. A Z-type (resp. Z-expression) represents an
H-type (resp. H-expression) with a single superimposed \emph{cursor}.

The only base cases in these inductive grammars are $\zwsel{\htau}$ and
$\zwsel{\hexp}$, which identify the H-type or H-expression that the cursor
is on. All of the other forms correspond to the recursive forms in the
syntax of H-types and H-expressions, and contain exactly one ``hatted''
subterm that identifies the subtree where the cursor will be found. Any
other sub-term is ``dotted'', i.e. it is either an H-type or
H-expression. Taken together, every Z-type and Z-expression contains
exactly one selected H-type or H-expression by construction. This can be
understood as an instance of Huet's \emph{zipper pattern}
\cite{JFP::Huet1997} (which, coincidentally, Huet encountered while
implementing a structure editor.)

We write $\removeSel{\ztau}$ for the H-type constructed by erasing the
cursor from $\ztau$, which we refer to as the \emph{cursor erasure} of
$\ztau$. This straightforward metafunction is defined as follows:
\begin{align*}
\removeSel{\zwsel{\htau}} & = \htau\\
\removeSel{\tarr{\ztau}{\htau}} & = \tarr{\removeSel{\ztau}}{\htau}\\
\removeSel{\tarr{\htau}{\ztau}} & = \tarr{\htau}{\removeSel{\ztau}}
\end{align*}

Similarly, we write $\removeSel{\zexp}$ for cursor erasure of $\zexp$. The definition of this metafunction is analogous, so we omit it
for concision.

This zipper structure is not the only way to model a cursor, though we have found it to be the most elegant for our present purposes. Another plausible strategy would be to formalize the notion of a relative path into an H-expression. This would then require defining the notion of consistency between a relative path and an H-expression, so we avoid it.

\subsection{Actions}\label{sec:actions}

We now arrive at the heart of Hazelnut: its \emph{bidirectional action
  semantics}.  Figure \ref{fig:action-syntax} defines the syntax of
\emph{actions}, $\alpha$, some of which involve \emph{directions},
$\delta$, and \emph{shapes}, $\psi$.

Expression actions are governed by two mutually defined judgements, 1) the
\emph{synthetic action judgement}:
\[
\performSyn{\hGamma}{\zexp}{\htau}{\alpha}{\zexp'}{\htau'}
\]
and 2) \emph{the analytic action judgement}:
\[
\performAna{\hGamma}{\zexp}{\htau}{\alpha}{\zexp'}
\]

In some Z-expressions, the cursor is in a type ascription, so we also need
a \emph{type action judgement}:
\[
\performTyp{\ztau}{\alpha}{\ztau'}
\]

\subsubsection{Sensibility}

These judgements are governed by a critical
metatheorem, \emph{action sensibility} (or simply \emph{sensibility}):
\begin{theorem}[Action Sensibility]
  \label{thrm:actsafe} ~
  \begin{enumerate}[itemsep=0px,partopsep=0px,topsep=0px]
  \item If $\hsyn{\hGamma}{\removeSel{\zexp}}{\htau}$ and
    $\performSyn{\hGamma}{\zexp}{\htau}{\alpha}{\zexp'}{\htau'}$ then
    $\hsyn{\hGamma}{\removeSel{\zexp'}}{\htau'}$.
  \item If $\hana{\hGamma}{\removeSel{\zexp}}{\htau}$ and
    $\performAna{\hGamma}{\zexp}{\htau}{\alpha}{\zexp'}$ then
    $\hana{\hGamma}{\removeSel{\zexp'}}{\htau}$.
  \end{enumerate}
\end{theorem}
\noindent In other words, if a Z-expression is
statically meaningful, i.e. its cursor erasure is well-typed, then
performing an action on it leaves the resulting Z-expression statically
meaningful. More specifically, the first clause of Theorem \ref{thrm:actsafe}
establishes that when an action is performed on a Z-expression whose cursor
erasure synthesizes an H-type, the result is a Z-expression whose cursor
erasure also synthesizes some (possibly different) H-type. The second
clause establishes that when an action is performed using the analytic
action judgement on an edit state whose cursor erasure analyzes against
some H-type, the result is a Z-expression whose cursor erasure also
analyzes against the same H-type.

This metatheorem deeply informs the design
of the rules, given starting in Sec. \ref{sec:action-subsumption}. Its
proof is by straightforward induction, so the
reader is encouraged to think about the relevant proof case when considering each action rule below.

No sensibility theorem is needed for the type action judgement because every syntactically well-formed type is meaningful in Hazelnut. (Adding type variables to the language would require defining both a type-level sensibility theorem and
type-level non-empty holes.)
\begin{figure}[t]
\hspace{-3px}$\arraycolsep=3pt\begin{array}{llcllll}
\mathsf{Action} & \alpha & ::= &
  \aMove{\delta} ~\vert~
  \aConstruct{\psi} ~\vert~
  \aDel ~\vert~
  \aFinish\\
\mathsf{Dir} & \delta & ::= &
  \dChildnm{n} ~\vert~
  \dParent\\
\mathsf{Shape} & \psi & ::= &
  \farr ~\vert~
  \fnum \\
& & \vert &
  \fasc ~\vert~
  \fvar{x} ~\vert~
  \flam{x} ~\vert~
  \fap ~\vert~
  \fnumlit{n} ~\vert~
  \fplus\\
& & \vert &
  {\color{gray}\fnehole}
\end{array}$
\caption{Syntax of actions.}
\label{fig:action-syntax}
\end{figure}

\subsubsection{Type Inconsistency}
In some of the rules below, we will need to supplement our definition of
type consistency from Figure \ref{fig:type-consistency} with a definition
of \emph{type inconsistency}, written $\tincompat{\htau}{\htau'}$. One can
define this notion either directly as the constructive negation of type
consistency, or as a separate inductively defined judgement with the following
key rule, which establishes that arrow types are inconsistent
with $\tnum$:
  \begin{equation*}
    \inferrule{ }{
      \tincompat{\tnum}{\tarr{\htau_1}{\htau_2}}
    }
  \end{equation*}
The mechanization proves that the judgemental definition of type
inconsistency is indeed the negation of type consistency.

\subsubsection{Action Subsumption}\label{sec:action-subsumption}

The action semantics includes a subsumption rule similar to the subsumption
rule, Rule (\ref{rule:ana-subsume}), in the statics:
\begin{equation}\label{rule:action-subsume}
  \inferrule{
    \hsyn{\hGamma}{\removeSel{\zexp}}{\htau'}\\
    \performSyn{\hGamma}{\zexp}{\htau'}{\alpha}{\zexp'}{\htau''}\\
    \tcompat{\htau}{\htau''}
  }{
    \performAna{\hGamma}{\zexp}{\htau}{\alpha}{\zexp'}
  }
\end{equation}
In other words, if the cursor erasure of the edit state synthesizes a type, $\htau'$,
then we defer to the synthetic action judgement. The cursor erasure of the Z-expression resulting from performing the action $\alpha$ synthetically could have a different type, $\htau''$, so we must check that it is consistent with the type provided for analysis, $\htau$.
The case for Rule (\ref{rule:action-subsume}) in the proof of Theorem
\ref{thrm:actsafe} goes through by induction and static subsumption,
i.e. Rule (\ref{rule:ana-subsume}). Algorithmically, subsumption should be
the rule of last resort (see Sec. \ref{sec:determinism} for further discussion.)

\subsubsection{Relative Movement}\label{sec:movement}
The rules below define relative movement within Z-types. They should be
self-explanatory:
\begin{subequations}
\begin{equation}
  \inferrule{ }{
    \performTyp{
      \zwsel{\tarr{\htau_1}{\htau_2}}
    }{
      \aMove{\dChildn{1}}
    }{
      \tarr{\zwsel{\htau_1}}{\htau_2}
    }
  }
\end{equation}
\begin{equation}\label{rule:move-arr-c2}
  \inferrule{ }{
    \performTyp{
      \zwsel{\tarr{\htau_1}{\htau_2}}
    }{
      \aMove{\dChildn{2}}
    }{
      \tarr{\htau_1}{\zwsel{\htau_2}}
    }
  }
\end{equation}
\begin{equation}\label{rule:move-parent-arr-left}
  \inferrule{ }{
    \performTyp{
      \tarr{\zwsel{\htau_1}}{\htau_2}
    }{
      \aMove{\dParent}
    }{
      \zwsel{\tarr{\htau_1}{\htau_2}}
    }
  }
\end{equation}
\begin{equation}\label{rule:move-parent-arr-right}
  \inferrule{ }{
    \performTyp{
      \tarr{{\htau_1}}{\zwsel{\htau_2}}
    }{
      \aMove{\dParent}
    }{
      \zwsel{\tarr{\htau_1}{\htau_2}}
    }
  }
\end{equation}
\end{subequations}
Two more rules are needed to recurse into the zipper structure. We define
these zipper rules in an action-independent manner in
Sec. \ref{sec:zipper-cases}.

The rules for relative movement within Z-expressions are similarly
straightforward. Movement is type-independent, so we defer to an auxiliary
expression movement judgement in both the analytic and synthetic case:
\begin{subequations}
\begin{equation}
\inferrule{
  \performMove{\zexp}{\aMove{\delta}}{\zexp'}
}{
  \performSyn{\hGamma}{\zexp}{\htau}{\aMove{\delta}}{\zexp'}{\htau}
}
\end{equation}
\begin{equation}
  \inferrule{
  \performMove{\zexp}{\aMove{\delta}}{\zexp'}
}{
  \performAna{\hGamma}{\zexp}{\htau}{\aMove{\delta}}{\zexp'}
}
\end{equation}
\end{subequations}
The expression movement judgement is defined as follows.

\paragraph{Ascription}
\begin{subequations}
  \begin{equation}
    \label{r:movefirstchild-asc}
  \inferrule{ }{
    \performTyp{
      \zwsel{\hexp : \htau}
    }{
      \aMove{\dChildn{1}}
    }{
      \zwsel{\hexp} : \htau
    }
  }
  \end{equation}
  \begin{equation}
    \label{r:movesecondchild-asc}
    \inferrule{ }{
    \performTyp{
      \zwsel{\hexp : \htau}
    }{
      \aMove{\dChildn{2}}
    }{
      \hexp : \zwsel{\htau}
    }
  }
\end{equation}
\begin{equation}
  \label{r:moveparent}
  \inferrule{ }{
    \performTyp{
      \zwsel{\hexp} : \htau
    }{
      \aMove{\dParent}
    }{
      \zwsel{\hexp : \htau}
    }
  }
\end{equation}
\begin{equation}\label{rule:move-parent-asc-right}
  \inferrule{ }{
    \performTyp{
      \hexp : \zwsel{\htau}
    }{
      \aMove{\dParent}
    }{
      \zwsel{\hexp : \htau}
    }
  }
\end{equation}

\paragraph{Lambda}\vspace{-3px}
\begin{equation}\label{r:movefirstchild-lam}
\inferrule{ }{
  \performMove{
    \zwsel{\hlam{x}{\hexp}}
  }{
    \aMove{\dChildn{1}}
  }{
    \hlam{x}{\zwsel{\hexp}}
  }
}
\end{equation}
\begin{equation}
  \inferrule{ }{
    \performMove{
      \hlam{x}{\zwsel{\hexp}}
    }{
      \aMove{\dParent}
    }{
      \zwsel{\hlam{x}{\hexp}}
    }
  }
\end{equation}
\paragraph{Application}\vspace{-5px}
\begin{equation}
  \inferrule{ }{
    \performMove{
      \zwsel{\hap{\hexp_1}{\hexp_2}}
    }{
      \aMove{\dChildn{1}}
    }{
      \hap{\zwsel{\hexp_1}}{\hexp_2}
    }
  }
\end{equation}
\begin{equation}
  \inferrule{ }{
    \performMove{
      \zwsel{\hap{\hexp_1}{\hexp_2}}
    }{
      \aMove{\dChildn{2}}
    }{
      \hap{\hexp_1}{\zwsel{\hexp_2}}
    }
  }
\end{equation}
\begin{equation}
  \inferrule{ }{
    \performMove{
      \hap{\zwsel{\hexp_1}}{\hexp_2}
    }{
      \aMove{\dParent}
    }{
      \zwsel{\hap{\hexp_1}{\hexp_2}}
    }
  }
\end{equation}
\begin{equation}\label{r:moveparent-ap2}
  \inferrule{ }{
    \performMove{
      \hap{{\hexp_1}}{\zwsel{\hexp_2}}
    }{
      \aMove{\dParent}
    }{
      \zwsel{\hap{\hexp_1}{\hexp_2}}
    }
  }
\end{equation}

\paragraph{Plus}
\begin{equation}
  \inferrule{ }{
    \performMove{
      \zwsel{\hadd{\hexp_1}{\hexp_2}}
    }{
      \aMove{\dChildn{1}}
    }{
      \hadd{\zwsel{\hexp_1}}{\hexp_2}
    }
  }
\end{equation}
\begin{equation}
  \inferrule{ }{
    \performMove{
      \zwsel{\hadd{\hexp_1}{\hexp_2}}
    }{
      \aMove{\dChildn{2}}
    }{
      \hadd{\hexp_1}{\zwsel{\hexp_2}}
    }
  }
\end{equation}
\begin{equation}
  \inferrule{ }{
    \performMove{
      \hadd{\zwsel{\hexp_1}}{\hexp_2}
    }{
      \aMove{\dParent}
    }{
      \zwsel{\hadd{\hexp_1}{\hexp_2}}
    }
  }
\end{equation}
\begin{equation}
  \inferrule{ }{
    \performMove{
      \hadd{{\hexp_1}}{\zwsel{\hexp_2}}
    }{
      \aMove{\dParent}
    }{
      \zwsel{\hadd{\hexp_1}{\hexp_2}}
    }
  }
\end{equation}

\paragraph{Non-Empty Hole}
\begin{equation}
\inferrule{ }{
  \performMove{
    \zwsel{\hhole{\hexp}}
  }{
    \aMove{\dChildn{1}}
  }{
    \hhole{\zwsel{\hexp}}
  }
}
\end{equation}
\begin{equation}\label{r:moveparent-hole}
  \inferrule{ }{
    \performMove{
      \hhole{\zwsel{\hexp}}
    }{
      \aMove{\dParent}
    }{
      \zwsel{\hhole{\hexp}}
    }
  }
\end{equation}
Again, additional rules are needed to recurse into the zipper structure,
but we will define these zipper rules in an action-independent manner in
Sec. \ref{sec:zipper-cases}.
\end{subequations}

The rules above are numerous and fairly uninteresting. That makes them
quite hazardous -- we might make a typo or forget a rule absent-mindedly. One check
against this is to establish that movement actions do not
change the cursor erasure, as in Theorem \ref{lemma:movement-erasure}.

\begin{theorem}[Movement Erasure Invariance]\label{lemma:movement-erasure} ~
  \begin{enumerate}[itemsep=0px,partopsep=0px,topsep=0px]
  \item If $\performMove{\ztau}{\aMove{\delta}}{\ztau'}$ then
    $\removeSel{\ztau}=\removeSel{\ztau'}$.

  \item If $\hsyn{\hGamma}{\removeSel{\zexp}}{\htau}$ and
    $\performSyn{\hGamma}{\zexp}{\htau}{\aMove{\delta}}{\zexp'}{\htau'}$
    then $\removeSel{\zexp}=\removeSel{\zexp'}$ and $\htau=\htau'$.

  \item If $\hana{\hGamma}{\removeSel{\zexp}}{\htau}$ and
    $\performAna{\hGamma}{\zexp}{\htau}{\aMove{\delta}}{\zexp'}$ then
    $\removeSel{\zexp}=\removeSel{\zexp'}$.

\end{enumerate}
\end{theorem}
\noindent Theorem \ref{lemma:movement-erasure} is useful also in that the
relevant cases of Theorem \ref{thrm:actsafe} are straightforward by its
application.

Another useful check is to establish \emph{reachability}, i.e. that it is
possible, through a sequence of movement actions, to move the cursor from
any position to any other position within a well-typed H-expression.

This requires developing machinery for reasoning about sequences of
actions. There are two possibilities: we can either add a sequencing
action, $\alpha; \alpha$, directly to the syntax of actions, or we can
define a syntax for lists of actions, $\bar{\alpha}$, together with
iterated action judgements. To keep the core of the action semantics small,
we take the latter approach in Figure \ref{fig:multistep}.

A simple auxiliary judgement, $\bar\alpha~\mathsf{movements}$ (not shown) establishes that $\bar\alpha$ consists only of actions of
the form $\aMove{\delta}$.

With these definitions, we can state reachability as follows:

\begin{theorem}[Reachability]\label{thrm:reachability} ~
  \begin{enumerate}[itemsep=0px,partopsep=0px,topsep=0px]
  \item If $\removeSel{\ztau}=\removeSel{\ztau'}$ then there exists some
    $\bar\alpha$ such that $\bar{\alpha}~\mathsf{movements}$ and
    $\performTypI{\ztau}{\bar\alpha}{\ztau'}$.

  \item If $\hsyn{\hGamma}{\removeSel{\zexp}}{\htau}$ and
    $\removeSel{\zexp}=\removeSel{\zexp'}$ then there exists some
    $\bar{\alpha}$ such that $\bar{\alpha}~\mathsf{movements}$ and
    $\performSynI{\hGamma}{\zexp}{\htau}{\bar\alpha}{\zexp'}{\htau}$.

  \item If $\hana{\hGamma}{\removeSel{\zexp}}{\htau}$ and
    $\removeSel{\zexp}=\removeSel{\zexp'}$ then there exists some
    $\bar{\alpha}$ such that $\bar{\alpha}~\mathsf{movements}$ and
    $\performAnaI{\hGamma}{\zexp}{\htau}{\bar{\alpha}}{\zexp'}$.
  \end{enumerate}
\end{theorem}

The simplest way to prove Theorem \ref{thrm:reachability} is to break it
into two lemmas. Lemma \ref{lemma:reach-up} establishes that you can always
move the cursor to the outermost position in an expression. This serves as
a check on our $\aMove{\dParent}$ rules.
\begin{lemma}[Reach Up]\label{lemma:reach-up} ~
  \begin{enumerate}[itemsep=0px,partopsep=0px,topsep=0px]
  \item If $\removeSel{\ztau}=\htau$ then there exists some $\bar\alpha$
    such that $\bar\alpha~\mathsf{movements}$ and
    $\performTypI{\ztau}{\bar\alpha}{\zwsel{\htau}}$.

  \item If $\hsyn{\hGamma}{\hexp}{\htau}$ and $\removeSel{\zexp}=\hexp$
    then there exists some $\bar\alpha$ such that
    $\bar\alpha~\mathsf{movements}$ and
    $\performSynI{\hGamma}{\zexp}{\htau}{\bar\alpha}{\zwsel{\hexp}}{\htau}$.

  \item If $\hana{\hGamma}{\hexp}{\htau}$ and $\removeSel{\zexp}=\hexp$
    then there exists some $\bar\alpha$ such that
    $\bar\alpha~\mathsf{movements}$ and
    $\performAnaI{\hGamma}{\zexp}{\htau}{\bar\alpha}{\zwsel{\hexp}}$.
  \end{enumerate}
\end{lemma}
Lemma \ref{lemma:reach-down} establishes that you can always move the
cursor from the outermost position to any other position. This serves as a
check on our $\aMove{\dChildnm{n}}$ rules.
\begin{lemma}[Reach Down]\label{lemma:reach-down} ~
  \begin{enumerate}[itemsep=0px,partopsep=0px,topsep=0px]
  \item If $\removeSel{\ztau}=\htau$ then there exists some $\bar\alpha$
    such that $\bar\alpha~\mathsf{movements}$ and
    $\performTypI{\zwsel{\htau}}{\bar\alpha}{\ztau}$.

  \item If $\hsyn{\hGamma}{\hexp}{\htau}$ and $\removeSel{\zexp}=\hexp$
    then there exists some $\bar\alpha$ such that
    $\bar\alpha~\mathsf{movements}$ and
    $\performSynI{\hGamma}{\zwsel{\hexp}}{\htau}{\bar\alpha}{\zexp}{\htau}$.

  \item If $\hana{\hGamma}{\hexp}{\htau}$ and $\removeSel{\zexp}=\hexp$
    then there exists some $\bar\alpha$ such that
    $\bar\alpha~\mathsf{movements}$ and
    $\performAnaI{\hGamma}{\zwsel{\hexp}}{\htau}{\bar\alpha}{\zexp}$.
  \end{enumerate}
\end{lemma}
Theorem \ref{thrm:reachability} follows by straightforward composition of
these two lemmas. The proofs we give of these three theorems in the
mechanization do not produce the shortest sequence of actions to witness
reachability, which would resemble something like a lowest common ancestor
computation. In future versions of Hazelnut that use the produced witnesses
for automatic tool support it may make sense to engineer these proofs
differently; here we are only concerned with whether the theorems are true.

\begin{figure}
$\mathsf{ActionList}$~~$\bar{\alpha} ::= \cdot ~\vert~ \alpha; \bar{\alpha}$\vspace{4px}\\
\fbox{$\performTypI{\ztau}{\bar{\alpha}}{\ztau'}$}

\vspace{-10px}\begin{subequations}
\begin{minipage}{0.35\linewidth}
\begin{equation}
\inferrule{ }{
    \performTypI{\ztau}{\cdot}{\ztau}
}
\end{equation}
\end{minipage}
\begin{minipage}{0.65\linewidth}
\begin{equation}
\inferrule{
  \performTyp{\ztau}{\alpha}{\ztau'}\\
  \performTypI{\ztau'}{\bar{\alpha}}{\ztau''}
}{
  \performTypI{\ztau}{\alpha; \bar{\alpha}}{\ztau''}
}
\end{equation}
\end{minipage}
\end{subequations}

\fbox{$\performSynI{\hGamma}{\zexp}{\htau}{\bar{\alpha}}{\zexp'}{\htau'}$}
\vspace{-10px}

\begin{subequations}\label{rules:iterated-syn}
\hspace{-12px}
\begin{mathpar}
\inferrule{ }{
  \performSynI{\hGamma}{\zexp}{\htau}{\cdot}{\zexp}{\htau}
}
~~~~~~
\inferrule{
  \performSyn{\hGamma}{\zexp}{\htau}{\alpha}{\zexp'}{\htau'}\\\\
  \performSynI{\hGamma}{\zexp'}{\htau'}{\bar{\alpha}}{\zexp''}{\htau''}
}{
  \performSynI{\hGamma}{\zexp}{\htau}{\alpha; \bar{\alpha}}{\zexp''}{\htau''}
}
~~~~~
\text{(\ref*{rules:iterated-syn}a-b)}
\end{mathpar}
\end{subequations}

\fbox{$\performAna{\hGamma}{\zexp}{\htau}{\bar{\alpha}}{\zexp'}$}
\vspace{-12px}
\begin{subequations}\label{rules:iterated-ana}
\begin{mathpar}
~~~~~~~\inferrule{ }{
  \performAnaI{\hGamma}{\zexp}{\htau}{\cdot}{\zexp}
}
~~~~~~~~~~~~~~
\inferrule{
  \performAna{\hGamma}{\zexp}{\htau}{\alpha}{\zexp'}\\\\
  \performAnaI{\hGamma}{\zexp'}{\htau}{\bar\alpha}{\zexp''}
}{
  \performAnaI{\hGamma}{\zexp}{\htau}{\alpha; \bar\alpha}{\zexp''}
}
~~~~~~~~~~
\text{(\ref*{rules:iterated-ana}a-b)}
\end{mathpar}
\end{subequations}
\caption{Iterated Action Judgements}
\label{fig:multistep}
\end{figure}
\subsubsection{Construction}\label{sec:construction} The construction
actions, $\aConstruct{\psi}$, are used to construct terms of a shape
indicated by $\psi$ at the cursor.

\paragraph{Types} The $\aConstruct{\farr}$ action constructs an arrow
type. The H-type under the cursor becomes the argument type, and the cursor
is placed on an empty return type hole:
\begin{subequations}
  \begin{equation}
    \label{r:contarr}
  \inferrule{ }{
    \performTyp{
      \zwsel{\htau}
    }{
      \aConstruct{\farr}
    }{
      \tarr{\htau}{\zwsel{\tehole}}
    }
  }
\end{equation}
This choice is formally arbitrary -- it would have also been sensible to use the type under the cursor as the return type, for example. Indeed, we could consider defining both of these using different shapes. We avoid this for the sake of simplicity.

The $\aConstruct{\fnum}$ action replaces an empty type hole under the
cursor with the $\tnum$ type:
  \begin{equation}
    \label{r:contnum}
  \inferrule{ }{
    \performTyp{
      \zwsel{\tehole}
    }{
      \aConstruct{\fnum}
    }{
      \zwsel{\tnum}
    }
  }
\end{equation}
\end{subequations}

\begin{subequations}

\paragraph{Ascription} The $\aConstruct{\fasc}$ action operates differently
depending on whether the H-expression under the cursor synthesizes a type
or is being analyzed against a type. In the first case, the synthesized
type appears in the ascription:
\begin{equation}
  \label{r:constructasc}
  \inferrule{ }{
    \performSyn{\hGamma}{\zwsel{\hexp}}{\htau}{\aConstruct{\fasc}}{\hexp : \zwsel{\htau}}{\htau}
  }
\end{equation}
In the second case, the type provided for analysis appears in the ascription:
\begin{equation}
  \inferrule{ }{
    \performAna{\hGamma}{\zwsel{\hexp}}{\htau}{\aConstruct{\fasc}}{\hexp : \zwsel{\htau}}
  }
\end{equation}

\paragraph{Variables} The $\aConstruct{\fvar{x}}$ action places the
variable $x$ into an empty hole. If that hole is being asked to synthesize
a type, then the result synthesizes the hypothesized type:
\begin{equation}
  \label{r:conevar}
  \inferrule{ }{
    \performSyn{\hGamma, x : \htau}{\zwsel{\hehole}}{\tehole}{\aConstruct{\fvar{x}}}{\zwsel{x}}{\htau}
  }
\end{equation}

If the hole is being analyzed against a type that is consistent with the
hypothesized type, then the action semantics goes through the {action
  subsumption rule} described in Sec. \ref{sec:action-subsumption}. If the
hole is being analyzed against a type that is inconsistent with the
hypothesized type, $x$ is placed inside a hole:
\begin{equation}
 \label{r:conevar2}
  \inferrule{
    \tincompat{\htau}{\htau'}
  }{
    \performAna{\hGamma, x : \htau'}{\zwsel{\hehole}}{\htau}{\aConstruct{\fvar{x}}}{\hhole{\zwsel{x}}}
  }
\end{equation}
The rule above featured on Line 16 of Figure \ref{fig:second-example}.

\paragraph{Lambdas} The $\aConstruct{\flam{x}}$ action places a lambda
abstraction binding $x$ into an empty hole. If the empty hole is being
asked to synthesize a type, then the result of the action is a lambda
ascribed the type $\tarr{\tehole}{\tehole}$, with the cursor on the
argument type hole (again, arbitrarily):
\begin{equation}
  \label{r:conelamhole}
  \inferrule{\alpha=\aConstruct{\flam{x}}}{
    \performSyn
      {\hGamma}
      {\zwsel{\hehole}}
      {\tehole}
      {\alpha}
      {\hlam{x}{\hehole} : \tarr{\zwsel{\tehole}}{\tehole}}
      {\tarr{\tehole}{\tehole}}
  }
\end{equation}
The type ascription is necessary because lambda expressions do not
synthesize a type. (Type-annotated function definitions often
arise as a single syntactic construct in full-scale languages like ML.)

If the empty hole is being analyzed against a type with
matched arrow type, then no ascription is necessary:
\begin{equation}\label{rule:performAna-lam-1}
  \inferrule{
    \arrmatch{\htau}{\tarr{\htau_1}{\htau_2}}
  }{
    \performAna
      {\hGamma}
      {\zwsel{\hehole}}
      {\htau}
      {\aConstruct{\flam{x}}}
      {\hlam{x}{\zwsel{\hehole}}}
  }
\end{equation}

Finally, if the empty hole is being analyzed against a type that has no
matched arrow type, expressed in the premise as inconsistency with
$\tarr{\tehole}{\tehole}$, then a lambda ascribed the type
$\tarr{\tehole}{\tehole}$ is inserted inside a hole, which defers the type
inconsistency as previously discussed:
\begin{equation}\label{rule:performAna-construct-lam-2}
  \inferrule{
    \tincompat{\htau}{\tarr{\tehole}{\tehole}}
  }{
    \performAna
      {\hGamma}
      {\zwsel{\hehole}}
      {\htau}
      {\aConstruct{\flam{x}}}
      {\hhole{
        \hlam{x}{\hehole} : \tarr{\zwsel{\tehole}}{\tehole}
      }}
  }
\end{equation}

\paragraph{Application} The $\aConstruct{\fap}$ action applies the
expression under the cursor. The following rule handles the case where the
synthesized type has matched function type:
\begin{equation}
  \label{r:coneapfn}
  \inferrule{
    \arrmatch{\htau}{\tarr{\htau_1}{\htau_2}}
  }{
    \performSyn
      {\hGamma}
      {\zwsel{\hexp}}
      {\htau}
      {\aConstruct{\fap}}
      {\hap{\hexp}{\zwsel{\hehole}}}
      {\htau_2}
  }
\end{equation}
If the expression under the cursor synthesizes a type that is inconsistent
with an arrow type, then we must place that expression inside a hole to
maintain Theorem \ref{sec:holes}:
\begin{equation}
  \inferrule{
    \tincompat{\htau}{\tarr{\tehole}{\tehole}}
  }{
    \performSyn
      {\hGamma}
      {\zwsel{\hexp}}
      {\htau}
      {\aConstruct{\fap}}
      {\hap{\hhole{\hexp}}{\zwsel{\hehole}}}
      {\tehole}
  }
\end{equation}


\paragraph{Numbers} The $\aConstruct{\fnumlit{n}}$ action replaces an empty
hole with the number expression $\hnum{n}$. If the empty hole is being
asked to synthesize a type, then the rule is straightforward:
\begin{equation}
  \label{r:conenumnum}
  \inferrule{ }{
    \performSyn
      {\hGamma}
      {\zwsel{\hehole}}
      {\tehole}
      {\aConstruct{\fnumlit{n}}}
      {\zwsel{\hnum{n}}}
      {\tnum}
  }
\end{equation}
If the empty hole is being analyzed against a type that is inconsistent
with $\tnum$, then we must place the number expression inside the hole:
\begin{equation}
  \inferrule{
    \tincompat{\htau}{\tnum}
  }{
    \performAna
      {\hGamma}
      {\zwsel{\hehole}}
      {\htau}
      {\aConstruct{\fnumlit{n}}}
      {\hhole{\zwsel{\hnum{n}}}}
  }
\end{equation}

The $\aConstruct{\fplus}$ action constructs a plus expression with the
expression under the cursor as its first argument (again, arbitrarily.) If that expression
synthesizes a type consistent with $\tnum$, then the rule is
straightforward:
\begin{equation}\label{rule:construct-plus-compat}
  \inferrule{
    \tcompat{\htau}{\tnum}
  }{
    \performSyn
      {\hGamma}
      {\zwsel{\hexp}}
      {\htau}
      {\aConstruct{\fplus}}
      {\hadd{\hexp}{\zwsel{\hehole}}}
      {\tnum}
  }
\end{equation}
Otherwise, we must place that expression inside a hole:
\begin{equation}
  \inferrule{
    \tincompat{\htau}{\tnum}
  }{
    \performSyn
      {\hGamma}
      {\zwsel{\hexp}}
      {\htau}
      {\aConstruct{\fplus}}
      {\hadd{\hhole{\hexp}}{\zwsel{\hehole}}}
      {\tnum}
  }
\end{equation}

\paragraph{Non-Empty Holes} The final shape is $\fnehole$. This explicitly
places the expression under the cursor inside a hole:
\begin{equation}
\inferrule{ }{
  \performSyn
    {\hGamma}
    {\zwsel{\hexp}}
    {\htau}
    {\aConstruct{\fnehole}}
    {\hhole{\zwsel{\hexp}}}
    {\tehole}
}
\end{equation}\end{subequations}

The $\fnehole$ shape is grayed out in Figure \ref{fig:action-syntax}
because we do not generally expect the programmer to perform it explicitly -- other
actions automatically insert holes when a type inconsistency would
arise.
The inclusion of this rule
simplifies the statement of the constructability theorem, discussed next.

\paragraph{Constructability}
To check that we have defined ``enough'' construct actions, we need to
establish that we can start from an empty hole and arrive at any well-typed
expression with the cursor on the outside. This simpler statement is
sufficient because Lemma \ref{lemma:reach-down} allows us to then move the
cursor anywhere else inside the constructed term. As with reachability, we
rely on the iterated action judgements defined in Figure
\ref{fig:multistep}.
\begin{theorem}[Constructability]\label{thrm:constructability} ~
  \begin{enumerate}[itemsep=0px,partopsep=0px,topsep=0px]
  \item For every $\htau$ there exists $\bar\alpha$ such that
    $\performTypI{\zwsel{\tehole}}{\bar\alpha}{\zwsel{\htau}}$.

  \item If $\hsyn{\hGamma}{\hexp}{\htau}$ then there exists $\bar\alpha$
    such
    that: $$\performSynI{\hGamma}{\zwsel{\hhole{}}}{\tehole}{\bar\alpha}{\zwsel{\hexp}}{\htau}$$

  \item If $\hana{\hGamma}{\hexp}{\htau}$ then there exists $\bar\alpha$
    such
    that: $$\performAnaI{\hGamma}{\zwsel{\hhole{}}}{\htau}{\bar\alpha}{\zwsel{\hexp}}$$
  \end{enumerate}
\end{theorem}
Without the $\fnehole$ shape, this theorem as stated would not hold. For example, it is not possible to construct well-typed H-expressions where non-empty holes appear superfluously without the $\fnehole$ shape. Note also that although none of the shapes that we have defined can be dropped without losing this theorem, some construction rules could be dropped. In particular, rules that insert non-empty holes automatically could be dropped because the $\aConstruct{\fnehole}$ action can always be used instead. We included them because we are interested in the mechanics of automatic non-empty hole insertion.

\subsubsection{Deletion} The $\aDel$ action inserts an empty hole at the
cursor, deleting what was there before.

The type action rule for $\aDel$ is self-explanatory:
\begin{equation}
  \inferrule{ }{
    \performTyp{
      \zwsel{\htau}
    }{
      \aDel
    }{
      \zwsel{\tehole}
    }
  }
\end{equation}

Deletion within a Z-expression is similarly straightforward:
\begin{subequations}
\begin{equation}
  \inferrule{ }{
    \performSyn{\hGamma}{\zwsel{\hexp}}{\htau}{\aDel}{\zwsel{\hehole}}{\tehole}
  }
\end{equation}
\begin{equation}
  \inferrule{ }{
    \performAna{\hGamma}{\zwsel{\hexp}}{\htau}{\aDel}{\zwsel{\hehole}}
  }
\end{equation}
\end{subequations}
Unlike the relative movement and construction actions, there is no
``checksum'' theorem for deletion. The rules do not inspect the structure
of the expression in the cursor, so they both match our intuition and will
be correct in any extension of the language without modification.

\subsubsection{Finishing}
The final action we need to consider is $\aFinish$, which finishes the
non-empty hole under the cursor.

If the non-empty hole appears in synthetic position, then it can always be
finished:
\begin{subequations}
  \begin{equation}
  \inferrule{
    \hsyn{\hGamma}{\hexp}{\htau'}
  }{
    \performSyn
      {\hGamma}
      {\zwsel{\hhole{\hexp}}}
      {\tehole}
      {\aFinish}
      {\zwsel{\hexp}}
      {\htau'}
  }
\end{equation}

If the non-empty hole appears in analytic position, then it can only be
finished if the type synthesized for the enveloped expression is consistent
with the type that the hole is being analyzed against. This amounts to
analyzing the enveloped expression against the provided type (by
subsumption):
\begin{equation}\label{r:finishana}
  \inferrule{
    \hana{\hGamma}{\hexp}{\htau}
  }{
    \performAna
      {\hGamma}
      {\zwsel{\hhole{\hexp}}}
      {\htau}
      {\aFinish}
      {\zwsel{\hexp}}
  }
\end{equation}
\end{subequations}
Like deletion, there is no need for a ``checksum'' theorem for the
finishing action.

\subsubsection{Zipper Cases}\label{sec:zipper-cases} The rules defined so
far handle the base cases, i.e. the cases where the action has ``reached''
the expression under the cursor. We also need to define the recursive
cases, which propagate the action into the subtree where the cursor
appears, as encoded by the zipper structure. For types, the zipper rules
are straightforward:
\begin{subequations}
\begin{equation}
  \inferrule{
    \performTyp{\ztau}{\alpha}{\ztau'}
  }{
    \performTyp{
      \tarr{\ztau}{\htau}
    }{
      \alpha
    }{
      \tarr{\ztau'}{\htau}
    }
  }
\end{equation}
  \begin{equation}
  \inferrule{
    \performTyp{\ztau}{\alpha}{\ztau'}
  }{
    \performTyp{
      \tarr{\htau}{\ztau}
    }{
      \alpha
    }{
      \tarr{\htau}{\ztau'}
    }
  }
\end{equation}
\end{subequations}
For expressions, the zipper rules essentially follow the structure of the
corresponding rules in the statics.

\begin{subequations}
In particular, when the cursor is in the body of a lambda expression, the
zipper case mirrors Rule (\ref{rule:syn-lam}):
\begin{equation}
\inferrule{
  \arrmatch{\htau}{\tarr{\htau_1}{\htau_2}}\\
  \performAna
    {\hGamma, x : \htau_1}
    {\zexp}
    {\htau_2}
    {\alpha}
    {\zexp'}
}{
  \performAna
    {\hGamma}
    {\hlam{x}{\zexp}}
    {\htau}
    {\alpha}
    {\hlam{x}{\zexp'}}
}
\end{equation}

When the cursor is in the function position of an application, the rule
mirrors Rule (\ref{rule:syn-ap}):
\begin{equation}
  \inferrule{
    \hsyn{\hGamma}{\removeSel{\zexp}}{\htau_2}\\
    \performSyn
      {\hGamma}
      {\zexp}
      {\htau_2}
      {\alpha}
      {\zexp'}
      {\htau_3}\\\\
    \arrmatch{\htau_3}{\tarr{\htau_4}{\htau_5}}\\
    \hana{\hGamma}{\hexp}{\htau_4}
  }{
    \performSyn
      {\hGamma}
      {\hap{\zexp}{\hexp}}
      {\htau_1}
      {\alpha}
      {\hap{\zexp'}{\hexp}}
      {\htau_5}
  }
\end{equation}

The situation is similar when the cursor is in argument position:
\begin{equation}
  \inferrule{
    \hsyn{\hGamma}{\hexp}{\htau_2}\\
    \arrmatch{\htau_2}{\tarr{\htau_3}{\htau_4}}\\
    \performAna
      {\hGamma}
      {\zexp}
      {\htau_3}
      {\alpha}
      {\zexp'}
  }{
    \performSyn
      {\hGamma}
      {\hap{\hexp}{\zexp}}
      {\htau_1}
      {\alpha}
      {\hap{\hexp}{\zexp'}}
      {\htau_4}
  }
\end{equation}

The rules for the addition operator mirror Rule (\ref{rule:syn-plus}):
\begin{equation}
  \inferrule{
    \performAna
      {\hGamma}
      {\zexp}
      {\tnum}
      {\alpha}
      {\zexp'}
  }{
    \performSyn
      {\hGamma}
      {\hadd{\zexp}{\hexp}}
      {\tnum}
      {\alpha}
      {\hadd{\zexp'}{\hexp}}
      {\tnum}
  }
\end{equation}
\begin{equation}
  \inferrule{
    \performAna
      {\hGamma}
      {\zexp}
      {\tnum}
      {\alpha}
      {\zexp'}
  }{
    \performSyn
      {\hGamma}
      {\hadd{\hexp}{\zexp}}
      {\tnum}
      {\alpha}
      {\hadd{\hexp}{\zexp'}}
      {\tnum}
  }
\end{equation}

When the cursor is in the expression position of an ascription, we use the
analytic  judgement, mirroring Rule (\ref{rule:syn-asc}):
\begin{equation}
\inferrule{
  \performAna
    {\hGamma}
    {\zexp}
    {\htau}
    {\alpha}
    {\zexp'}
}{
  \performSyn
    {\hGamma}
    {\zexp : \htau}
    {\htau}
    {\alpha}
    {\zexp' : \htau}
    {\htau}
}
\end{equation}

When the cursor is in the type position of an ascription, we must re-check
the ascribed expression because the cursor erasure might have changed (in
practice, one would optimize this check to only occur if the cursor erasure
did change):
\begin{equation}\label{rule:zipper-asc}
\inferrule{
  \performTyp{\ztau}{\alpha}{\ztau'}\\
  \hana{\hGamma}{\hexp}{\removeSel{\ztau'}}
}{
  \performSyn
    {\hGamma}
    {\hexp : \ztau}
    {\removeSel{\ztau}}
    {\alpha}
    {\hexp : \ztau'}
    {\removeSel{\ztau'}}
}
\end{equation}

Finally, if the cursor is inside a non-empty hole, the relevant zipper rule
mirrors Rule (\ref{rule:syn-ehole}):
\begin{equation}
  \inferrule{
    \hsyn{\hGamma}{\removeSel{\zexp}}{\htau}\\
    \performSyn
      {\hGamma}
      {\zexp}
      {\htau}
      {\alpha}
      {\zexp'}
      {\htau'}\\
  }{
    \performSyn
      {\hGamma}
      {\hhole{\zexp}}
      {\tehole}
      {\alpha}
      {\hhole{\zexp'}}
      {\tehole}
  }
\end{equation}

Theorem \ref{thrm:actsafe} directly checks the correctness of these
rules. Moreover, the zipper rules arise ubiquitously in derivations of edit
steps, so the proofs of the other ``check'' theorems, e.g. Reachability and
Constructability, serve as a check that none of these rules have been
missed.
\end{subequations}

\subsection{Determinism}\label{sec:determinism}
A last useful property to consider is \emph{action determinism}, i.e. that
performing an action produces a unique result. The following theorem establishes determinism for type actions:
\begin{theorem}[Type Action Determinism]
\label{thrm:type-actdet} If $\performTyp{\ztau}{\alpha}{\ztau'}$ and
    $\performTyp{\ztau}{\alpha}{\ztau''}$ then $\ztau'=\ztau''$.
\end{theorem}

The corresponding theorem for expression actions would be stated as follows:
  \begin{enumerate}[itemsep=0px,partopsep=0px,topsep=0px]
  \item If $\hsyn{\hGamma}{\removeSel{\zexp}}{\htau}$ and
    $\performSyn{\hGamma}{\zexp}{\htau}{\alpha}{\zexp'}{\htau'}$ and
    $\performSyn{\hGamma}{\zexp}{\htau}{\alpha}{\zexp''}{\htau''}$ then
    $\zexp' = \zexp''$ and $\htau' = \htau''$.
  \item If $\hana{\hGamma}{\removeSel{\zexp}}{\htau}$ and
    $\performAna{\hGamma}{\zexp}{\htau}{\alpha}{\zexp'}$ and
    $\performAna{\hGamma}{\zexp}{\htau}{\alpha}{\zexp''}$ then $\zexp' =
    \zexp''$.
  \end{enumerate}

This is not a theorem of the system as described so far. The reason
is somewhat subtle: two construction actions,
$\aConstruct{\fasc}$ and $\aConstruct{\flam{x}}$, behave differently in the
analytic case than they do in the synthetic case. The problem is that both rules can
``fire'' when considering a Z-expression in analytic position due to action subsumption. This is a technically valid but ``morally'' invalid use of action subsumption: subsumption is included in the system to be used as a rule of last resort, i.e. it should only be applied when no other analytic action rule can fire.

There are several possible ways to address this problem. One approach would be to modify the judgement forms to internalize this notion of ``rule of last resort''. This approach is related to focusing from proof theory \cite{Simmons11tr}. However, this approach would substantially complicate our presentation of the system.

The approach that we take leaves the system unchanged. Instead, we define predicates over \emph{derivations} of the expression action judgements that exclude those derivations that apply subsumption prematurely, i.e. when another rule could have been applied. We call such derivations \emph{subsumption-minimal derivations}. We can establish determinism for subsumption-minimal derivations.
\begin{theorem}[Expression Action Determinism] ~
  \begin{enumerate}[itemsep=0px,partopsep=0px,topsep=0px]
  \item If $\hsyn{\hGamma}{\removeSel{\zexp}}{\htau}$ and
    $\mathcal{D}_1 : \performSyn{\hGamma}{\zexp}{\htau}{\alpha}{\zexp'}{\htau'}$ and
    $\mathcal{D}_2 : \performSyn{\hGamma}{\zexp}{\htau}{\alpha}{\zexp''}{\htau''}$ and $\subminsyn{\mathcal{D}_1}$ and $\subminsyn{\mathcal{D}_2}$ then
    $\zexp' = \zexp''$ and $\htau' = \htau''$.
  \item If $\hana{\hGamma}{\removeSel{\zexp}}{\htau}$ and
    $\mathcal{D}_1 : \performAna{\hGamma}{\zexp}{\htau}{\alpha}{\zexp'}$ and
    $\mathcal{D}_2 : \performAna{\hGamma}{\zexp}{\htau}{\alpha}{\zexp''}$ and $\subminana{\mathcal{D}_1}$ and $\subminana{\mathcal{D}_2}$ then $\zexp' =
    \zexp''$.
  \end{enumerate}
\end{theorem}

The mechanization, discussed next, defines the predicates $\subminsyn{\mathcal{D}}$ and $\subminana{\mathcal{D}}$. In addition, it defines a mapping from
any derivation into a corresponding subsumption-minimal derivation. Implementations of Hazelnut need only implement this subsumption-minimal fragment.

\subsection{Mechanization}
\label{sec:mech}\label{sec:mt}

In order to
formally establish that our design meets our stated objectives, we have
mechanized the semantics and metatheory of Hazelnut as described above using the Agda proof
assistant \cite{norell:thesis} (also see the Agda Wiki, hosted
at \url{http://wiki.portal.chalmers.se/agda/}.) This development is available in the supplemental material. The mechanization also includes the  extension to Hazelnut described in Sec. \ref{sec:extending}.

The documentation includes a more detailed discussion of the technical
representation decisions that we made. The main idea is standard: we encode
each judgement as a dependent type. The rules defining the judgements
become the constructors of this type, and derivations are terms of these
type. This is a rich setting that allows proofs to take advantage of
pattern matching on the shape of derivations, closely matching standard
on-paper proofs. No proof automation was used, because the proof structure
itself is likely to be interesting to researchers who plan to build upon
our work.

We adopt Barendregt's convention for bound
variables \cite{urban}. Hazelnut's semantics does not need substitution, so
we do not need to adopt more sophisticated encodings
(e.g. \cite{lh09unibind,Pouillard11}.)

\section{Extending Hazelnut}\label{sec:extending}
\newcounter{sumtypedef}
\renewcommand{\thesumtypedef}{\theequation\alph{sumtypedef}}

\newcommand{\Define}[1]{(\refstepcounter{sumtypedef}\thesumtypedef\label{#1})}

In this section, we will conservatively extend Hazelnut with binary {sum
types} to demonstrate how the rules and the rich metatheory developed in
the previous section serve to guide and constrain this and other such
efforts.

\paragraph{Syntax.}
The first step is to extend the syntax of H-types and H-expressions
with the familiar forms \cite{pfpl}:\vspace{-2px}
\begin{grammar}
$\mathsf{HTyp}$ & $\htau$ & \bnfas & $\cdots \bnfalt \tsum{\htau}{\htau}$
\\
$\mathsf{HExp}$ & $\hexp$ & \bnfas & $\cdots
\bnfalt \hinj{i}{\hexp}
\bnfalt \hcase{\hexp}{x}{\hexp}{y}{\hexp}$
\end{grammar}\vspace{-2px}
Recall that binary sum types introduce a new type-level connective,
$\tsum{\htau_1}{\htau_2}$. The introductory forms are the \emph{injections},
$\hinj{i}{\hexp}$; here, we consider only binary sums, so $i\in\{\mathsf{L}, \mathsf{R}\}$.
The elimination form is case
analysis,~$\hcase{\hexp}{x}{\hexp_1}{y}{\hexp_2}$.

\begin{figure}
{
\judgbox{\tcompat{\htau_1}{\htau_2}}{
}
\vspace{-14px}\begin{equation}
  \inferrule{
    \tcompat{\htau_1}{\htau_1'}
    \\
    \tcompat{\htau_2}{\htau_2'}
    }
   {\tcompat{\tsum{\htau_1}{\htau_2}}{\htau_1' + \htau_2'}}
\end{equation}

\judgbox{\sumhasmatched{\htau}{\htau_1+\htau_2}}{$\htau$ has matched sum type $\htau_1+\htau_2$}
\begin{subequations}
\begin{minipage}{.448\linewidth}
\begin{equation}
\inferrule{ }
{\sumhasmatched{\hehole}{\hehole + \hehole}}
\end{equation}
\end{minipage}
\begin{minipage}{.55\linewidth}
\begin{equation}
\inferrule{ }
{\sumhasmatched{\tsum{\htau_1}{\htau_2}}{\tsum{\htau_1}{\htau_2}}}
\end{equation}
\end{minipage}
\end{subequations}

\vspace{3px}
\judgbox{\hana{\hGamma}{\hexp}{\htau}}{
}
\begin{subequations}
\begin{equation}
\inferrule{
  \sumhasmatched{\htau_{+}}{\tsum{\htau_{\mathsf{L}}}{\htau_\mathsf{R}}}\\
  \hana{ \hGamma }{ \hexp }{ \htau_i }
}
{ \hana{ \hGamma }{ \hinj{i}{\hexp} }{ \htau_{+} } }~\text{$(i \in \{\mathsf{L}, \mathsf{R}\})$}
\end{equation}
\begin{equation}
\inferrule
{ \hsyn{ \hGamma }{ \hexp }{ \htau_{+} }
  \\
  \sumhasmatched{\htau_{+}}{\tsum{\htau_1}{\htau_2}}
  \\\\
  \hana{ \hGamma, x:\htau_1 }{ \hexp_1 }{ \htau }
  \\
  \hana{ \hGamma, y:\htau_2 }{ \hexp_2 }{ \htau }
}
{ \hana{ \hGamma }{ \hcase{\hexp}{x}{\hexp_1}{y}{\hexp_2} }{ \htau } }
\end{equation}
\end{subequations}
\vspace{-4px}
\caption{The statics of sums.}
\label{fig:sum-statics}
\vspace{8px}
\judgbox{\performTyp{\ztau}{\alpha}{\ztau'}}{
}\vspace{-19px}
\begin{subequations}
\begin{equation}\label{rule:construct-sum}
  \inferrule{ }
{
  \performTyp{\zwsel{\htau}}{\aConstruct{\fsum}}
             {\tsum{{\htau}}{\zwsel{\hehole}}}
}
\end{equation}
\noindent\begin{minipage}{.5\linewidth}
 \vspace{-8px}\begin{equation}\label{rule:zipper-sum-left}
  \inferrule{
    \performTyp{\ztau}{\alpha}{\ztau'}
  }{
    \performTyp{
      \tsum{\ztau}{\htau}
    }{
      \alpha
    }{
      \tsum{\ztau'}{\htau}
    }
  }
\end{equation}\end{minipage}\begin{minipage}{.5\linewidth}
\vspace{-8px}\begin{equation}\label{rule:zipper-sum-right}
  \inferrule{
    \performTyp{\ztau}{\alpha}{\ztau'}
  }{
    \performTyp{
      \tsum{\htau}{\ztau}
    }{
      \alpha
    }{
      \tsum{\htau}{\ztau'}
    }
  }
\end{equation}
\end{minipage}
\end{subequations}
\judgbox{\performAna{\hGamma}{\zexp}{\htau}{\alpha}{\zexp'}}
{
}
\begin{subequations}
\begin{equation}\label{rule:performAna-inj-1}
  \inferrule{ \sumhasmatched{\htau_{+}}{\tsum{\htau_{1}}{\htau_{2}}} }
  {
  \performAna{\hGamma}{\zwsel{\hehole}}
              {\htau_{+}}
              {\aConstruct{\finj{i}}}
              {\hinj{i}{\zwsel{\hehole}}}
  }
\end{equation}
\begin{equation}\label{rule:performAna-inj-2}
  \inferrule{ \tincompat{\htau}{ \tsum{\hehole}{\hehole} } }
        {
  \performAna{\hGamma}{\zwsel{\hehole}}
              {\htau}
              {\aConstruct{\finj{i}}}
              {\hhole{
                  \hinj{i}{\hehole}
                  : \tsum{\zwsel{\hehole}}{\hehole}
              }}
        }
\end{equation}
\begin{equation}\label{rule:performAna-case}
  \inferrule{ }{
  \performAna{\hGamma}{\zwsel{\hehole}}
              {\htau}
              {\aConstruct{\fcase{x}{y}}}
              {\hcase{\zwsel{\hehole}}{x}{\hehole}{y}{\hehole}}
  }
\end{equation}
\begin{equation}\label{rule:zipper-inj}
\inferrule
{
  \sumhasmatched{\htau_{+}}{\tsum{\htau_{\mathsf{L}}}{\htau_{\mathsf{R}}}}\\
  \performAna{\hGamma}{\zexp}{\htau_i}{\alpha}{\zexp'}
}{
  \performAna
  {\hGamma}
  {\hinj{i}{\zexp }}
  {\htau_{+}}
  {\alpha}
  {\hinj{i}{\zexp' }}
}~\text{$(i\in\{\mathsf{L}, \mathsf{R}\})$}
\end{equation}
\begin{equation}\label{rule:zipper-case-1}
\inferrule
{
  \hsyn
  {\hGamma}
  {\removeSel{\zexp}}
  {\htau_0}
  \\
  \performSyn
  {\hGamma}
  {\zexp}
  {\htau_0}
  {\alpha}
  {\zexp'}
  {\htau_{+}}
  \\
  \sumhasmatched{\htau_{+}}{\tsum{\htau_1}{\htau_2}}
  \\\\
  \hana{\hGamma, x:\htau_1}{\hexp_1}{\htau}
  \\
  \hana{\hGamma, y:\htau_2}{\hexp_2}{\htau}
}{
  \performAna
  {\hGamma}
  {\hcase{\zexp }{x}{\hexp_1}{y}{\hexp_2}}
  {\htau}
  {\alpha}
  {\hcase{\zexp'}{x}{\hexp_1}{y}{\hexp_2}}
}
\end{equation}
\begin{equation}\label{rule:zipper-case-2}
\inferrule
{
  \hsyn{\hGamma}{\hexp}{\htau_{+}}
  \\
  \sumhasmatched{\htau_{+}}{\tsum{\htau_1}{\htau_2}}
  \\
  \performAna{\hGamma, x:\htau_1}{\zexp_1}{\htau}{\alpha}{\zexp_1'}
}{
  \performAna
  {\hGamma}
  {\hcase{\hexp}{x}{\zexp_1}{y}{\hexp_2}}
  {\htau}
  {\alpha}
  {\hcase{\hexp}{x}{\zexp_1'}{y}{\hexp_2}}
}
\end{equation}
\begin{equation}\label{rule:zipper-case-3}
\inferrule
{
  \hsyn{\hGamma}{\hexp}{\htau_{+}}
  \\
  \sumhasmatched{\htau_{+}}{\tsum{\htau_1}{\htau_2}}
  \\
  \performAna{\hGamma, y:\htau_2}{\zexp_2}{\htau}{\alpha}{\zexp_2'}
}{
  \performAna
  {\hGamma}
  {\hcase{\hexp}{x}{\hexp_1}{y}{\zexp_2}}
  {\htau}
  {\alpha}
  {\hcase{\hexp}{x}{\hexp_1}{y}{\zexp_2'}}
}
\end{equation}
\end{subequations}
\judgbox{\performSyn{\hGamma}{\zexp}{\htau}{\alpha}{\zexp'}{\htau'}}
{
}\vspace{-5px}
\begin{subequations}
\begin{equation}\label{rule:performSyn-inj}
\inferrule
{
  \alpha = \aConstruct{\finj{i}}
}{
  \performSyn
  {\hGamma}
  {\zwsel{\hhole{}}}
  {\htau}
  {\alpha}
  {\hinj{i}{{\hhole{}}} : \tsum{\zwsel{\tehole}}{\tehole}}
  {\tehole + \tehole}
}
\end{equation}
\begin{equation}\label{rule:performSyn-case-1}
\inferrule{
  \alpha = \aConstruct{\fcase{x}{y}}\\
  \sumhasmatched{\htau}{\tsum{\htau_1}{\htau_2}}
}{
  \performSyn
  {\hGamma}
  {\zwsel{\hexp}}
  {\htau}
  {\alpha}
  {\hcase{\hexp}{x}{\zwsel{\hhole{}}}{y}{\hhole{}} : \tehole}
  {\tehole}
}
\end{equation}
\begin{equation}\label{rule:performSyn-case-2}
\inferrule{
  \alpha = \aConstruct{\fcase{x}{y}}\\
  \tincompat{\htau}{\tsum{\tehole}{\tehole}}
}{
  \performSyn
  {\hGamma}
  {\zwsel{\hexp}}
  {\htau}
  {\alpha}
  {\hcase{\hhole{\zwsel{\hexp}}}{x}{{\hhole{}}}{y}{\hhole{}} : \tehole}
  {\tehole}
}
\end{equation}
\end{subequations}
\vspace{-3px}
\caption{The construction \& zipper action rules for sums.}
\label{fig:sum-action}
}
\end{figure}

\begin{figure}
\judgbox{\performMove{\ztau}{\aMove{\delta}}{\ztau'}}{\hspace{0.6\linewidth}(25a-d)}
\begin{displaymath}
\begin{array}{@{}rcl}
  \TABperformMove
      {\zwsel{\tsum{\htau_1}{\htau_2}}}
      {\aMove{\dChildn{1}}}
      {      {\tsum{\zwsel{\htau_1}}{\htau_2}}}
  \\
  \TABperformMove
      {\zwsel{\tsum{\htau_1}{\htau_2}}}
      {\aMove{\dChildn{2}}}
      {      {\tsum{\htau_1}{\zwsel{\htau_2}}}}
  \\
  \TABperformMove
      {      {\tsum{\zwsel{\htau_1}}{\htau_2}}}
      {\aMove{\dParent}}
      {\zwsel{\tsum{{\htau_1}}{{\htau_2}}}}
  \\
  \TABperformMove
      {      {\tsum{{\htau_1}}{{\zwsel{\htau_2}}}}}
      {\aMove{\dParent}}
      {\zwsel{\tsum{{\htau_1}}{\htau_2}}}
  \\[2mm]
\end{array}
\end{displaymath}
\judgbox{\performMove{\zexp}{\aMove{\delta}}{\zexp'}}{\hspace{0.61\linewidth}(26a-h)}
\begin{displaymath}
\begin{array}{@{}rcl}
  \TABperformMove
      {\zwsel{\hinj{i}{\hexp}}}
      {\aMove{\dChildn{1}}}
      {\hinj{i}{\zwsel{\hexp}}}
  \\
  \TABperformMove
      {\hinj{i}{\zwsel{\hexp}}}
      {\aMove{\dParent}}
      {\zwsel{\hinj{i}{\hexp}}}
  \\[2mm]
  \TABperformMove
      {\zwsel{\hcase{\hexp}{x}{\hexp_1}{y}{\hexp_2}}}
      {\aMove{\dChildn{1}}}
      {      {\hcase{\zwsel{\hexp}}{x}{\hexp_1}{y}{\hexp_2}}}
  \\
  \TABperformMove
      {\zwsel{\hcase{\hexp}{x}{\hexp_1}{y}{\hexp_2}}}
      {\aMove{\dChildn{2}}}
      {      {\hcase{\hexp}{x}{\zwsel{\hexp_1}}{y}{\hexp_2}}}
  \\
  \TABperformMove
      {\zwsel{\hcase{\hexp}{x}{\hexp_1}{y}{\hexp_2}}}
      {\aMove{\dChildn{3}}}
      {      {\hcase{\hexp}{x}{\hexp_1}{y}{\zwsel{\hexp_2}}}}
  \\
  \TABperformMove
      {      {\hcase{\zwsel{\hexp}}{x}{\hexp_1}{y}{\hexp_2}}}
      {\aMove{\dParent}}
      {\zwsel{\hcase{{\hexp}}{x}{\hexp_1}{y}{\hexp_2}}}
  \\
  \TABperformMove
      {      {\hcase{{\hexp}}{x}{\zwsel{\hexp_1}}{y}{\hexp_2}}}
      {\aMove{\dParent}}
      {\zwsel{\hcase{{\hexp}}{x}{\hexp_1}{y}{\hexp_2}}}
  \\
  \TABperformMove
      {      {\hcase{{\hexp}}{x}{\hexp_1}{y}{\zwsel{\hexp_2}}}}
      {\aMove{\dParent}}
      {\zwsel{\hcase{{\hexp}}{x}{\hexp_1}{y}{\hexp_2}}}
\end{array}
\end{displaymath}
\vspace{-3px}
\caption{Movement actions for sums.}
\label{fig:sum-move}
\end{figure}

Next, we must correspondingly extend the syntax of Z-types and
Z-expressions, following Huet's zipper pattern \cite{JFP::Huet1997}:\iftr \vspace{-11px} \else \vspace{-3px} \fi 
\begin{grammar}
$\mathsf{ZTyp}$ & $\ztau$ & \bnfas & $\cdots \bnfalt \tsum{\ztau}{\htau} \bnfalt \tsum{\htau}{\ztau}$
\\
$\mathsf{ZExp}$ & $\zexp$ & \bnfas & $\cdots
\bnfalt \hinj{i}{\zexp}
\bnfalt \hcase{\zexp}{x}{\hexp}{y}{\hexp}$
\\
&& $\bnfalt$ & $\hcase{\hexp}{x}{\zexp}{y}{\hexp}
\bnfalt\hcase{\hexp}{x}{\hexp}{y}{\zexp}$
\end{grammar}
Notice that for each H-type or H-expression form of arity $n$, there are
$n$ corresponding Z-type or Z-expression forms, each of which has a single
``hatted'' subterm. The remaining subterms are ``dotted''. We must also
extend the definition of cursor erasure, e.g. for types:
\begin{align*}
\removeSel{\tsum{\ztau}{\htau}} & = \tsum{\removeSel{\ztau}}{\htau}\\
\removeSel{\tsum{\htau}{\ztau}} & = \tsum{\htau}{\removeSel{\ztau}}
\end{align*}
The rules for Z-expressions are analogous:
\begin{align*}
\removeSel{\hinj{i}{\zexp}} & = \hinj{i}{\removeSel{\zexp}}\\
\removeSel{\hcase{\zexp}{x}{\hexp_1}{y}{\hexp_2}} & = \hcase{\removeSel{\zexp}}{x}{\hexp_1}{y}{\hexp_2}\\
\removeSel{\hcase{\hexp}{x}{\zexp_1}{y}{\hexp_2}} & = \hcase{\hexp}{x}{\removeSel{\zexp_1}}{y}{\hexp_2}\\
\removeSel{\hcase{\hexp}{x}{\hexp_1}{y}{\zexp_2}} & = \hcase{\hexp}{x}{\hexp_1}{y}{\removeSel{\zexp_2}}
\end{align*}

Finally, we must extend the syntax of shapes:
\begin{grammar}
$\mathsf{Shape}$ & $\psi$ & \bnfas & $\cdots \bnfalt \fsum \bnfalt \finj{i} \bnfalt \fcase{x}{y}$
\end{grammar}
Notice that for each H-type or H-expression form, there is a corresponding
shape. The injection form had a formal parameter, $i$, so the corresponding
shape takes a corresponding argument (like $\fnumlit{n}$.) The case form
included two variable binders, so the corresponding shape takes two
variable arguments (like $\flam{x}$.)

\paragraph{Statics.}
We can now move on to the static semantics.

First, we must extend the type consistency relation as shown in
Figure \ref{fig:sum-statics}, following the example of covariant type
consistency rule for arrow types in
Figure \ref{fig:type-consistency}. Similarly, we need a notion of
a \emph{matched sum type} analogous to the notion of a matched
arrow type defined in Figure \ref{fig:type-consistency}.

The type analysis rules shown in Figure \ref{fig:sum-statics} are
essentially standard, differing only in that instead of immediately
requiring that a type be of the form $\tsum{\htau_1}{\htau_2}$, we use the
matched sum type judgement. We combine the two injection rules for
concision and define only a type analysis rule for the case form for
simplicity (see \cite{DBLP:conf/popl/CiminiS16} for additional machinery
that would be necessary for a synthetic rule.)

\paragraph{Action Semantics.}
Figures~\ref{fig:sum-action} and \ref{fig:sum-move} extend Hazelnut's
action semantics to support bidirectionally typed structure editing with sums.

Rule (\ref{rule:construct-sum}), the construction rule for sum types, and
Rules (\ref{rule:zipper-sum-left})-(\ref{rule:zipper-sum-right}), the
zipper rules for sum types, follow the corresponding rules for arrow
types. Were we to have missed any of these, the first clause of
Theorem \ref{thrm:constructability}, i.e. Constructability, would not be
conserved.

Rule (\ref{rule:performAna-inj-1}) constructs an injection when the type
provided for analysis has a matched sum type. This is analogous to Rule
(\ref{rule:performAna-lam-1}) for lambdas. Rule
(\ref{rule:performAna-inj-2}) constructs an injection when the type
provided for analysis is not consistent with sum types. This is analogous
to Rule (\ref{rule:performAna-construct-lam-2}) for lambdas. Rule
(\ref{rule:performAna-case}) is a straightforward rule for constructing
case expressions in empty holes. Rules
(\ref{rule:zipper-inj})-(\ref{rule:zipper-case-3}) are the zipper cases,
which follow the structure of the statics. Finally, we also define a single
new synthetic action rule, Rule (\ref{rule:performSyn-inj}), which allows
for the construction of an injection in synthetic position, with automatic
insertion of an ascription. This is analogous to Rule
(\ref{r:conelamhole}). If we had defined any of these rules incorrectly,
the Sensibility Theorem (Theorem \ref{thrm:actsafe}) would not be
conserved. Had we forgotten the analytic rules, the Constructability
Theorem (Theorem \ref{thrm:constructability}) would not be conserved.

Figure \ref{fig:sum-move} gives the relevant movement axioms. For concision
and clarity, we write these axioms in tabular form. Had we made a mistake
in any of these rules, the Movement Erasure Invariance theorem
(Theorem \ref{lemma:movement-erasure}) would not be conserved. Had we
forgotten any of these rules, the Reachability Theorem
(Theorem \ref{thrm:reachability}) would not be conserved.

\section{Implementation}
\label{sec:impl}
\subsection{Implementation Concepts}
\begin{figure}
\centering
\includegraphics[width=\columnwidth]{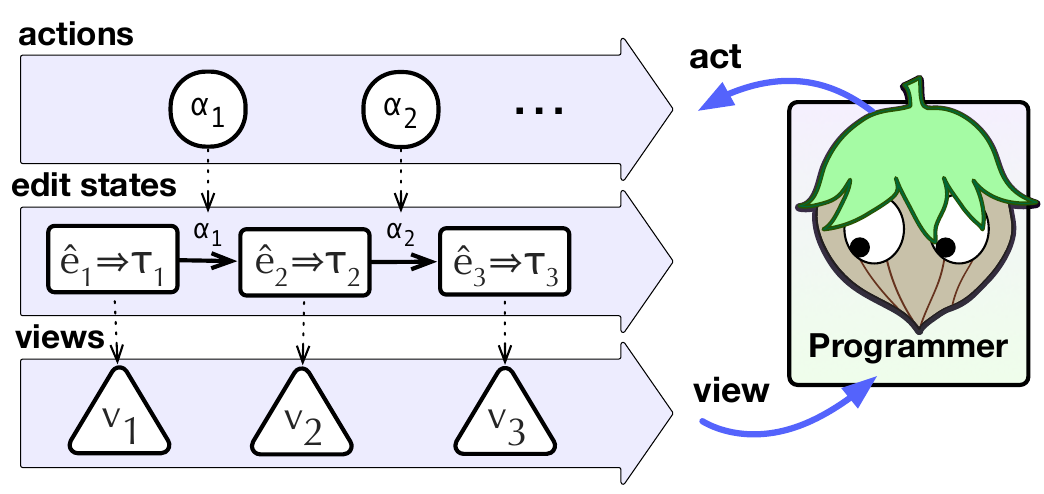}
\caption{Implementation Concepts}
\label{fig:impl-overview}
\end{figure}

Central to any implementation of Hazelnut is a stream of edit states whose
cursor erasures synthesize types under an empty context according to the
synthetic action judgement,
$\performSyn{\emptyset}{\zexp}{\htau}{\alpha}{\zexp'}{\htau'}$. The middle
row of Figure \ref{fig:impl-overview} diagrams this stream of edit
states. For example, the reader is encouraged to re-examine the
examples in Figure \ref{fig:first-example} and \ref{fig:second-example} --
the cursor erasure of each edit state synthesizes a type.

Because Theorem \ref{thrm:actsafe} expresses an invariant, the editor does not need to
typecheck the edit state anew on each action (though in some of the rules, e.g. Rule
(\ref{rule:zipper-asc}) which handles the situation where the type in a
type ascription changes, portions of the program would need to be
typechecked again.) In other words, many of the problems related to incrementality 
are simply not relevant. Storing the types of subtrees would allow for further
optimizations (e.g. of the zipper rules.)

The programmer examines a view generated from each edit state and produces
actions in some implementation-defined manner (e.g. using a keyboard,
mouse, touchscreen, voice interface, or neural interface), as diagrammed in Figure \ref{fig:impl-overview}. Each new action
causes a new abstract edit state to arise according to an implementation of
the action semantics. This then causes a new view to arise. This is a
simple event-based functional reactive programming model
\cite{Wan:2000:FRP:349299.349331}.

If an action is not well-defined according to Hazelnut's action semantics,
the implementation must reject it. In fact, the implementation is
encouraged to present an ``action palette'' that either hides or visibly
disables actions that are not well-defined (see below.)

\subsection{HZ}
We have developed a simple reference implementation, HZ, of Hazelnut extended with sum types as described in Sec. \ref{sec:extending}.  In
order to reach a wide audience, we decided to implement HZ in the web
browser.  To take advantage of a mature implementation of the FRP
model, we chose to implement HZ using
OCaml\footnote{\url{https://ocaml.org/}}, the OCaml React
library\footnote{\url{http://erratique.ch/software/react}}, and the \texttt{js\_of\_ocaml}
compiler and associated libraries
\cite{DBLP:conf/ml/Balat06}\footnote{\url{http://ocsigen.org/js\_of\_ocaml/}}. 

The core semantics have
been implemented in a functional style that follows the presentation in this paper closely.

The view computation renders the model (i.e. a Z-expression paired with a type) as nested HTML \texttt{div} elements matching the tree structure of the corresponding Z-expression. This tree is stylized separately using CSS. The action palette is a collection of buttons and text boxes which
are disabled when the corresponding action cannot be performed. We determine this by simply attempting to perform the
action internally and handling the exception that is raised when an action
is undefined. (An optimization that is not in HZ would be to implement a version of the action semantics that simply computes a boolean, rather than the resulting edit state, for the purposes of action validation.) Each action form also has a corresponding keyboard shortcut. For actions that take arguments, the keyboard shortcut moves the cursor into a text box. The action validation occurs on every change to the text box.

This implementation is not, of course, meant to score marks for
usability or performance (though both are surprisingly good for such a simple system.) Instead, as a simple reference implementation, it allows the reader to interact with the system presented in this paper, to aid in their understanding.  Additionally, we expect its source code to be of use to others who are interested in
layering a more fluid user interface atop the core semantics, or extensions thereof.

\section{Related Work and Discussion}\label{sec:rw}
\subsection{Structure Editors}
Syntactic structure editors have a long history -- the Cornell Program
Synthesizer~\cite{teitelbaum_cornell_1981} was first introduced in
1981. Novice programmers have been a common target for structure
editors. For example, GNOME~\cite{garlan_gnome:_1984} was developed to
teach programming to undergraduates.
Alice~\cite{Conway:2000:ALL:332040.332481} is a 3-D programming language
with an integrated structure editor for teaching novice CS undergraduate
students. Scratch~\cite{Resnick:2009:SP:1592761.1592779} is a structure
editor targeted at children ages 8 to 16.
TouchDevelop \cite{tillmann_touchdevelop:_2011} incorporates a structure
editor for programming on touch-based devices, and is used to teach high
school students. An implementation of Hazelnut might be useful in teaching
students about the typed lambda calculus, though that has not been our
explicit aim with this work.

Not all structure editors are for educational purposes. For example,
mbeddr \cite{voelter_mbeddr:_2012} is a structure editor for a C-based
programming language (nominally, for programming embedded systems.)
Lamdu~\cite{lamdu}, like Hazelnut, is a structure editor for a statically
typed functional language. It is designed for use by professional
programmers.

The examples given so far either do not attempt to reason statically about
types and binding, or do not attempt to maintain well-typedness as an edit
invariant. This can pose problems, for reasons discussed in Sec. \ref{sec:introduction}. One apparent exception is Unison~\cite{unison}, a structure
editor for a typed functional language similar to Haskell. Like Hazelnut,
it seems to define some notion of well-typedness for expressions with holes
(though there is no technical documentation on virtually any aspect of its
design.) Unlike Hazelnut, it does not have a notion analogous to Hazelnut's
notion of a non-empty hole. As such, programmers must construct programs in
a rigid outside-in manner, as discussed in Sec. \ref{sec:example}. Another
system with the same problem is Polymorphic Blocks, a block-based user
interface where the structure of block connectors encodes a
type \cite{DBLP:conf/chi/LernerFG15}.

We fundamentally differ from these projects in our design philosophy: we
consider it essential to start by building type theoretic foundations,
which are independent of nearly all decisions about the user interface (other than our choice to use an explicit cursor.) In
contrast, these editors have developed innovative user interfaces (e.g. see
the discussion in \cite{DBLP:conf/sle/VolterSBK14}) but lack a principled
foundational calculus. In this respect, we follow the philosophical
approach taken by languages that are rooted in the type theoretic tradition
and have gone to great effort to develop a clear metatheory, like Standard
ML \cite{mthm97-for-dart,Harper00atype-theoretic,Lee:2007:TMM:1190216.1190245}.  In the future, we hope
that these lines of research will merge to produce a human-usable typed
structure editor with sound formal foundations. Our contribution, then, is
in defining and analyzing the theoretical properties of a small
foundational calculus that could be extended to achieve this vision.
Our implementation resembles the minimal structure editor defined in
Haskell by Sufrin and De Moor \cite{sufrin1999modeless}.

Some structure editor \emph{generators} do operate on formal or semi-formal
definitions of an underlying language. For example, the Synthesizer
Generator~\cite{Reps:1984:SG:390010.808247} allows the user to define an
attribute grammar-based language implementation that then can be used to
generate a structured editor. CENTAUR~\cite{Borras:1988:CS:64140.65005}
produces a language specific environment from a user defined formal
specification of a language. Barista is a programmatic toolkit for building
structure editors \cite{ko_barista:_2006}. mbeddr is built on top of the
commercial JetBrains MPS framework for constructing structure
editors \cite{voelter2011language,DBLP:journals/software/VoelterWK15}. These
systems do not give a semantics to the edit states of the structure editor
itself, or maintain non-trivial edit invariants, as Hazelnut does.

Related to structure editors are value editors, which operate directly on
simple values (but not generally expressions or functions) of a programming
language. For example, Eros is a typed value editor based in
Haskell \cite{DBLP:conf/icfp/Elliott07}.

Other work has attempted to integrate structure editing features into
text editors. For example, recent work has used syntactic placeholders
reminiscent of our expression holes to decrease the percentage of
edit states that are malformed \cite{Amorim:2016:PSC:2997364.2997374}. This work does not consider the semantics of placeholders.

Prior work has also explored
formal definitions of text editor commands, e.g. using functional
combinators \cite{DBLP:journals/scp/Sufrin82}.

\subsection{Gradual Type Systems}
A significant contribution of this paper is the discovery of a clear
technical relationship between typed structure editing and gradual
typing. In particular, the machinery necessary to give a reasonable
semantics to type holes -- i.e. type consistency and type matching --
coincides with that developed in the study of gradual type systems for
functional languages. The pioneering work of Siek and Taha \cite{Siek06a}
introduced type consistency. Subsequent work developed the concept of type
matching \cite{DBLP:conf/popl/RastogiCH12,DBLP:conf/popl/GarciaC15} and has further
studied the notion of type consistency \cite{Garcia:2016:AGT:2837614.2837670}. In
retrospect, this relationship is perhaps unsurprising: gradual typing is,
notionally, also motivated by the idea of iterated development of a program
where every intermediate state is well-defined in some sense, albeit at
different granularity.

Recent work has discovered a systematic procedure for generating a
``gradual version'' of a standard type
system \cite{DBLP:conf/popl/CiminiS16}. This system, called the
Gradualizer, operates on a logic program that specifies a simple type
assignment system with some additional annotations to generate a
corresponding specification of a gradual type system. The authors leave
support for working with bidirectional type systems as future work. This
suggests the possibility of an analogous ``Editorializer'' that generates a
specification of a typed structure editor from a simple language
definition. Our exposition in Sec. \ref{sec:extending} certainly suggests
that many of the necessary definitions follow seemingly mechanically from
the definition of the static semantics, and the relationship with gradual
typing suggests that many of the technical details of this transformation
may already exist in the Gradualizer. One possibility we have explored
informally is to use Agda's reflection features to implement such a system.

An aspect of gradual typing that we did not touch on directly here is its
concern with assigning a dynamics to programs where type information is not
known, by inserting dynamic type casts \cite{Siek06a} or deducing evidence for consistency during evaluation \cite{Garcia:2016:AGT:2837614.2837670}. This would correspond to assigning
a dynamics to Hazelnut expressions with type holes such that a run-time
error occurs when a type hole is found to be unfillable through evaluation. This
may be useful as an exploratory programming tool.

\subsection{Bidirectional Type Systems}
Hazelnut is bidirectionally typed \cite{Pierce:2000:LTI:345099.345100,DBLP:conf/icfp/DaviesP00,DBLP:conf/tldi/ChlipalaPH05,bidi-tutorial,odersky2001colored}. Bidirectional type systems are notable in
that they are easy to define, easy to implement, produce simple error messages and support advanced language features \cite{dunfield2013complete}. For example, Scala \cite{odersky2001colored} and Agda \cite{norell:thesis} are both fundamentally bidirectionally typed languages.

\subsection{Type Reconstruction}
An alternative approach to type inference is to use a unification-based type reconstruction system, as in functional languages like ML and Haskell \cite{damas1982principal}. This is difficult to reconcile with the approach presented in this paper, because edit actions could introduce new  unification constraints that would require placing non-empty holes around terms far from the cursor.  A whole-program hole insertion pass after each edit action could perhaps be used to recover invariants similar to those presented here, but we leave the details as future work. Our contention is that a bidirectional approach is a sweet spot in the design space of interactive systems like Hazelnut because it precludes
``spooky errors at a distance''. Instead, the interaction is a sort of local
dialog between the programmer and the system involving simple, familiar
concepts -- types with holes -- rather than sets of constraints.

An intermediate approach would be to
layer unification-based type generation features atop the bidirectional system. This would amount to  interpreting type holes as unification
variables. For a simple calculus,
e.g. the STLC upon which Hazelnut is based, type inference for complete expressions is known to be
decidable, so type holes could be instantiated automatically once the expression that they appear within has been constructed. It would also be possible to flag expressions for which there does not exist any way to fill the type holes. In more complex settings, e.g. in a
dependently typed language, a partial decision procedure may still be
useful in this regard, both at edit-time and (just prior to)
run-time. Indeed, text editor modes for dependently typed proof assistants, e.g. for Agda,
attempt to do exactly this for indicated ``type holes'' (and do not always
succeed.)

\subsection{Exceptions}
Expression holes can be interpreted in several
ways. One straightforward interpretation is to treat them
like expressions that raise exceptions. Indeed,
placing \textt{raise Unimplemented} or similar in portions of an expression
that are under construction is a common practice across programming
languages today. The GHC dialect of Haskell recently introduced an explicit notion of a
typed hole that behaves similarly \cite{GHC/holes}.

\subsection{Type-Directed Program Synthesis}
Some text editor modes, e.g. those for proof assistants like
Agda \cite{norell:thesis} and Idris \cite{brady2013idris}, support a more
explicit hole-based programming model where indicated expression holes are
treated as sites where the system can be asked to automatically generate an
expression of an appropriate type. 

The Graphite system borrowed Eclipse's heuristic model of typed holes for Java to allow
library providers to associate interactive code generation interfaces with
types \cite{Omar:2012:ACC:2337223.2337324}.

More generally, the topic of type-directed program synthesis an active area
of research, e.g. \cite{DBLP:conf/pldi/OseraZ15}. By maintaining static
well-definedness throughout the editing process, Hazelnut provides
researchers interested in editor-integrated type-directed program synthesis
with a formal foundation upon which to build.

\subsection{Tactics}

Interactive proof refinement systems, e.g. those in LCF \cite{Gordon:1978:MIP:512760.512773}, and more recent typed tactic systems, e.g. Mtac for Coq \cite{ziliani2015mtac},
support an explicit model of a ``current'' typed hole that serves as the
target of program synthesis. Hazelnut differs in that edits can occur anywhere within a term.

A related approach is to interpret expression holes as the \emph{metavariables} of contextual modal type theory
(CMTT) \cite{DBLP:journals/tocl/NanevskiPP08}. In particular, expression
holes have types and are surrounded by contexts, just as metavariables in
CMTT are associated with types and contexts. This begins to clarify the
logical meaning of a typing derivation in Hazelnut -- it conveys
well-typedness relative to an (implicit) modal context that extracts each
expression hole's type and context. The modal context must be emptied --
i.e. the expression holes must be instantiated with expressions of the
proper type in the proper context -- before the expression can be
considered complete. This corresponds to the notion of modal necessity in
contextual modal logic.

We did not make the modal context explicit in our semantics because interactive program editing is not
merely hole filling in Hazelnut (i.e. the cursor need not be on a
hole.) Moreover, the hole's type and context become apparent as our action
semantics traverses the zipper structure on each step. For interactive
proof assistants that support a tactic model based directly on hole
filling, as just discussed, the connection to CMTT and similar systems is more useful. For
example, Beluga \cite{DBLP:conf/flops/Pientka10} is based on dependent CMTT
and aspects of Idris' editor support \cite{brady2013idris} are based on
McBride's OLEG \cite{mcbride2000dependently} and Lee and Friedman have
explored a lambda calculus with contexts for a similar
purpose \cite{DBLP:conf/icfp/LeeF96}.

One interesting avenue of future work is to elaborate expression holes to
CMTT's closures, i.e. CMTT terms of the form
$\mathsf{clo}(u; \text{id}(\Gamma))$ where $u$ is a unique metavariable
associated with each hole and $\text{id}(\Gamma)$ is the explicit identity
substitution. This would allow us to evaluate expressions with holes such
that the closure ``accumulates'' substitutions explicitly. When evaluation
gets ``stuck'' (as it can, for CMTT does not define a dynamics equipped
with a notion of progress under a non-empty modal context), it would then
be possible for the programmer to choose a hole from the visible
holes (which may have been duplicated) to edit in their original
context. Once finished, the CMTT hole instantiation operation, together
with a metatheorem that establishes that reduction commutes with
instantiation, would enable an ``edit and resume'' feature with a clear
formal basis. This notion of reduction commuting with instantiation has
also been studied in other
calculi \cite{DBLP:journals/entcs/Sands97}. Being able to edit a running
program also has connections to less formal work on ``live programming''
interfaces \cite{burckhardt2013s,lamdu}.

\section{Conclusion}
\label{sec:future}
This paper presented Hazelnut, a type theoretic structure editor
calculus. Our aim was to take a principled approach to its design by
formally defining its static semantics as well as its action semantics and
developing a rich metatheory. Moreover, we have mechanized substantial
portions of the metatheory, including the crucial Sensibility theorem that
establishes that every edit state is statically meaningful.

In addition to simplifying the job of an editor designer, typed structure
editors also promise to increase the speed of development by eliminating
redundant syntax and supporting higher-level primitive actions. However, we
did not discuss such ``edit costs'' here, because they depend on particular
implementation details, e.g. whether a keyboard or a mouse is in
use. Indeed, we consider it a virtue of this work that such implementation
details do not enter into our design.

\subsection{Future Work}
\subsubsection{Richer Languages}
Hazelnut is, obviously, a very limited language at its core. So the most
obvious avenue for future work is to increase the expressive power of this
language by extension. Our plan is to simultaneously maintain a
mechanization and implementation (following, for example, Standard ML \cite{Lee:2007:TMM:1190216.1190245}) as
we proceed, ultimately producing the first large-scale, formally verified
bidirectionally typed structure editor.

It is interesting to note that the demarcation between the language and the
editor is fuzzy (indeed, non-existent) in Hazelnut. There may well be
interesting opportunities in language design when the language is being
codesigned with a typed structure editor. It may be that certain language
features are unnecessary given a sufficiently advanced type-aware structure
editor (e.g. SML's \texttt{open}?), while other features may only be
practical with editor support. We intend to use Hazelnut and derivative
systems thereof as a platform for rigorously exploring such questions.

\subsubsection{Evaluation Strategies: A High-Dimensional Space}
The related work brought up in the previous section suggests three different evaluation strategies in
the presence of type holes:
\begin{enumerate}[noitemsep]
\item ...as preventing evaluation (the standard approach.)
\item ...as unknown types, in the gradual sense.
\item ...as unification variables.
\end{enumerate}
In addition, we have discussed four different evaluation strategies in the
presence of expression holes:
\begin{enumerate}[noitemsep]
\item ...as preventing evaluation (the standard approach.)
\item ...as causing exceptions.
\item ...as sites for automatic program synthesis.
\item ...as the closures of CMTT.
\end{enumerate}

Every combination of these choices could well be considered in the design
of a full-scale programming system in the spirit of Hazelnut. Indeed, the
user might be given several options from among these combinations,
depending on their usage scenario. Many of these warrant further inquiry.

\subsubsection{Editor Services}
There are various aspects of the editor model that we have not yet
formalized. For example, our action model does not consider how actions are
actually entered using, for example, key combinations or chords. In
practice, we would want also to suggest sensible compound actions, and
to rank these suggested actions in some reasonable
manner (perhaps based on usage data gathered from other users or code
repositories.) Designing a action suggestion semantics and a rigorous typed probability model over actions is one avenue of research that we have started to
explore, with intriguing initial results.

\subsubsection{Programmable Actions}
Our language of actions is intentionally primitive. However, even now it
acts much like a simple imperative command language. This suggests future
expansion to, for example, a true \emph{action macro} language, whereby
functional programs could themselves be lifted to the level of actions to
compute non-trivial compound actions. Such compound actions would give a
uniform description of transformations ranging from the simple -- like
``move the cursor to the next hole to the right'' -- to quite complex whole
program refactoring, while remaining subject to the core Hazelnut
metatheory. Techniques like those in advanced tactic systems, e.g. Mtac, might be useful
in proving these action macros correct \cite{ziliani2015mtac}.

\subsubsection{Views}
Another research direction is in exploring how types can be used to control
the presentation of expressions in the editor. For example, following an
approach developed in a textual setting of developing \emph{type-specific
languages} (TSLs), it should be possible to have the type that an
expression is being analyzed against define alternative display forms and
interaction modes \cite{TSLs}.

It should also be possible to develop the concept of semantic comments,
i.e. comments that mention semantic structures or even contain values. These would be subject to
the same operations, e.g. renaming, as other structures, helping to address
the problem of comments becoming inconsistent with code. This system, generalized sufficiently,
could one day help unify document editing with program editing.

\subsubsection{Collaborative Programming}
Finally, we did not consider any aspects of \emph{collaborative
programming}, such as a packaging system, a differencing algorithm for use
in a source control system, support for multiple simultaneous cursors for
different users, and so on. These are all interesting avenues for future
work.

\subsubsection{Empirical Evaluation}
Although we make few empirical claims in this paper, it is ultimately an
empirical question as to whether structure editors, and typed structure
editors, are practical. We hope to conduct user studies once a richer
semantics and a practical implementation thereof has been developed.

\subsubsection{More Theory}
Connections with gradual type systems and CMTT, discussed in the previous
section, seem likely to continue to be revealing.

The notion of having one of many possible locations within a term under a
cursor has a very strong intuitive connection to the proof theoretic notion
of focusing \cite{Simmons11tr}. Building closer connections with proof
theory (and category theory) is likely to be a fruitful avenue of further
inquiry.

\begin{quote}
\emph{In any case, these are but steps toward more graphical program-description
systems, for we will not forever stay confined to mere strings of symbols.}

--- Marvin Minsky, Turing Award lecture \cite{DBLP:journals/jacm/Minsky70}
\end{quote}

\acks

The authors would like to thank the anonymous referees at POPL and TFP 2016  for useful feedback on earlier drafts of this paper; Ed Morehouse and Carlo Angiuli for counsel on mechanization and Barendrecht's
Convention; Vincent Zeng for \textit{pro bono} artistic services; and YoungSeok Yoon for reanalyzing the data from \cite{6883030}.
This work was partially funded through a gift from Mozilla; the NSF grant \#CCF-1619282, 1553741 and 1439957; by AFRL and DARPA under agreement \#FA8750-16-2-0042; and by NSA lablet contract \#H98230-14-C-0140. 

\balance
\bibliographystyle{abbrvnat}
\bibliography{bibliography}

\iftr
\clearpage
\onecolumn
\appendix

\section{Hazelnut}
The full collection of rules defining the semantics of Hazelnut (not including the rules for the extension described in Sec. \ref{sec:extending}) are
reproduced here in their definitional order for reference. The rule names coincide with the corresponding
constructors in the Agda mechanization.
\subsection{H-Types and H-Expressions}
\subsubsection{Type Compatibility and Incompatibility}

\noindent\fbox{$\tcompat{\htau}{\htau'}$}
\begin{subequations}
  \begin{equation*}
    \tag{\rname{TCRefl}}
    \inferrule{ }{
      \tcompat{\htau}{\htau}
    }
  \end{equation*}
  \gap{}
  \begin{equation*}
    \tag{\rname{TCHole1}}
    \inferrule{ }{
      \tcompat{\htau}{\tehole}
    }
  \end{equation*}
  \gap{}
  \begin{equation*}
    \tag{\rname{TCHole2}}
    \inferrule{ }{
      \tcompat{\tehole}{\htau}
    }
  \end{equation*}
  \gap{}
  \begin{equation*}
    \tag{\rname{TCArr}}
    \inferrule{
      \tcompat{\htau_1}{\htau_1'}\\
      \tcompat{\htau_2}{\htau_2'}
    }{
      \tcompat{\tarr{\htau_1}{\htau_2}}{\tarr{\htau_1'}{\htau_2'}}
    }
  \end{equation*}
\end{subequations}

\noindent\fbox{$\tincompat{\htau}{\htau'}$}
\begin{subequations}
  \begin{equation*}
    \tag{\rname{ICNumArr1}}
    \inferrule{ }{
      \tincompat{\tnum}{\tarr{\htau_1}{\htau_2}}
    }
  \end{equation*}
  \gap{}
  \begin{equation*}
    \tag{\rname{ICNumArr2}}
    \inferrule{ }{
      \tincompat{\tarr{\htau_1}{\htau_2}}{\tnum}
    }
  \end{equation*}
  \gap{}
  \begin{equation*}
    \tag{\rname{ICArr1}}
    \inferrule{
      \tincompat{\htau_1}{\htau_3}
    }{
      \tincompat{\tarr{\htau_1}{\htau_2}}{\tarr{\htau_3}{\htau_4}}
    }
  \end{equation*}
  \gap{}
  \begin{equation*}
    \tag{\rname{ICArr2}}
    \inferrule{
      \tincompat{\htau_2}{\htau_4}
    }{
      \tincompat{\tarr{\htau_1}{\htau_2}}{\tarr{\htau_3}{\htau_4}}
    }
  \end{equation*}
\end{subequations}

\subsubsection{Function Type Matching}
\noindent
\fbox{$\arrmatch{\htau}{\tarr{\htau_1}{\htau_2}}$}
\begin{subequations}
  \begin{equation*}
    \tag{\rname{MAHole}}
    \inferrule{ }{
      \arrmatch{\tehole}{\tarr{\tehole}{\tehole}}
    }
  \end{equation*}
  \gap{}
  \begin{equation*}
    \tag{\rname{MAArr}}
    \inferrule{ }{
      \arrmatch{\tarr{\htau_1}{\htau_2}}{\tarr{\htau_1}{\htau_2}}
    }
  \end{equation*}
\end{subequations}

\subsubsection{Synthesis and Analysis}
The judgements $\hsyn{\hGamma}{\hexp}{\htau}$ and
$\hana{\hGamma}{\hexp}{\htau}$ are defined mutually inductively.

\noindent\fbox{$\hsyn{\hGamma}{\hexp}{\htau}$}
\begin{subequations}\label{Arules:hsyn}
  \begin{equation*}
    \tag{\rname{SAsc}}
    \inferrule{
      \hana{\hGamma}{\hexp}{\htau}
    }{
      \hsyn{\hGamma}{\hexp : \htau}{\htau}
    }
  \end{equation*}
  \gap{}
  \begin{equation*}
    \tag{\rname{SVar}}
    \inferrule{ }{
      \hsyn{\hGamma, x : \htau}{x}{\htau}
    }
  \end{equation*}
  \gap{}
  \begin{equation*}
    \tag{\rname{SAp}}
    \inferrule{
      \hsyn{\hGamma}{\hexp_1}{\htau}\\
      \arrmatch{\htau}{\tarr{\htau_2}{\htau'}}\\
      \hana{\hGamma}{\hexp_2}{\htau_2}
    }{
      \hsyn{\hGamma}{\hap{\hexp_1}{\hexp_2}}{\htau'}
    }
  \end{equation*}
  \gap{}
  \begin{equation*}
    \tag{\rname{SNum}}
    \inferrule{ }{
      \hsyn{\hGamma}{\hnum{n}}{\tnum}
    }
  \end{equation*}
  \gap{}
  \begin{equation*}
    \tag{\rname{SPlus}}
    \inferrule{
      \hana{\hGamma}{\hexp_1}{\tnum}\\
      \hana{\hGamma}{\hexp_2}{\tnum}
    }{
      \hsyn{\hGamma}{\hadd{\hexp_1}{\hexp_2}}{\tnum}
    }
  \end{equation*}
  \gap{}
  \begin{equation*}
    \tag{\rname{SHole}}
    \inferrule{ }{
      \hsyn{\hGamma}{\hehole}{\tehole}
    }
  \end{equation*}
  \gap{}
  \begin{equation*}
    \tag{\rname{SNEHole}}
    \inferrule{
      \hsyn{\hGamma}{\hexp}{\htau}
    }{
      \hsyn{\hGamma}{\hhole{\hexp}}{\tehole}
    }
  \end{equation*}
\end{subequations}
\noindent\fbox{$\hana{\hGamma}{\hexp}{\htau}$}
\begin{subequations}\label{Arules:hana}
  \begin{equation*}
    \tag{\rname{ASubsume}}
    \inferrule{
      \hsyn{\hGamma}{\hexp}{\htau'}\\
      \tcompat{\htau}{\htau'}
    }{
      \hana{\hGamma}{\hexp}{\htau}
    }
  \end{equation*}
  \gap{}
  \begin{equation*}
    \tag{\rname{ALam}}
    \inferrule{
      \arrmatch{\htau}{\tarr{\htau_1}{\htau_2}}\\
      \hana{\hGamma, x : \htau_1}{\hexp}{\htau_2}
    }{
      \hana{\hGamma}{\hlam{x}{\hexp}}{\htau}
    }
  \end{equation*}
\end{subequations}

\subsection{Z-Types and Z-Expressions}
\subsubsection{Type Cursor Erasure}
\noindent\fbox{$\removeSel{\ztau}=\htau$} is a metafunction defined as follows:
\begin{subequations}
  \begin{align}
    \tag{\rname{ETTop}}
    \removeSel{\zwsel{\htau}} & = \htau\\
    \tag{\rname{ETArrL}}
    \removeSel{\tarr{\ztau}{\htau}} & = \tarr{\removeSel{\ztau}}{\htau}\\
    \tag{\rname{ETArrR}}
    \removeSel{\tarr{\htau}{\ztau}} & = \tarr{\htau}{\removeSel{\ztau}}
  \end{align}
\end{subequations}

\subsubsection{Expression Cursor Erasure}
\noindent\fbox{$\removeSel{\zexp}=\hexp$} is a metafunction defined as follows:
\begin{subequations}
  \begin{align}
    \tag{\rname{EETop}}
    \removeSel{\zwsel{\hexp}} & = \hexp\\
    \tag{\rname{EEAscL}}
    \removeSel{\zexp : \htau} & = \removeSel{\zexp} : \htau\\
    \tag{\rname{EEAscR}}
    \removeSel{\hexp : \ztau} & = \hexp : \removeSel{\ztau}\\
    \tag{\rname{EELam}}
    \removeSel{\hlam{x}{\zexp}} & = \hlam{x}{\removeSel{\zexp}}\\
    \tag{\rname{EEApL}}
    \removeSel{\hap{\zexp}{\hexp}} & = \hap{\removeSel{\zexp}}{\hexp}\\
    \tag{\rname{EEApR}}
    \removeSel{\hap{\hexp}{\zexp}} & = \hap{\hexp}{\removeSel{\zexp}}\\
    \tag{\rname{EEPlusL}}
    \removeSel{\hadd{\zexp}{\hexp}} & = \hadd{\removeSel{\zexp}}{\hexp}\\
    \tag{\rname{EEPlusR}}
    \removeSel{\hadd{\hexp}{\zexp}} & = \hadd{\hexp}{\removeSel{\zexp}}\\
    \tag{\rname{EENEHole}}
    \removeSel{\hhole{\zexp}} &= \hhole{\removeSel{\zexp}}
  \end{align}
\end{subequations}

\subsection{Action Model}
\subsubsection{Type Actions}
\noindent\fbox{$\performTyp{\ztau}{\alpha}{\ztau'}$}
\paragraph{Type Movement}
\begin{subequations}
  \begin{equation*}
    \tag{\rname{TMArrChild1}}
    \inferrule{ }{
      \performTyp{
        \zwsel{\tarr{\htau_1}{\htau_2}}
      }{
        \aMove{\dChildn{1}}
      }{
        \tarr{\zwsel{\htau_1}}{\htau_2}
      }
    }
  \end{equation*}
  \gap{}
  \begin{equation*}
    \tag{\rname{TMArrChild2}}
    \inferrule{ }{
      \performTyp{
        \zwsel{\tarr{\htau_1}{\htau_2}}
      }{
        \aMove{\dChildn{2}}
      }{
        \tarr{\htau_1}{\zwsel{\htau_2}}
      }
    }
  \end{equation*}
  \gap{}
  \begin{equation*}
    \tag{\rname{TMArrParent1}}
    \inferrule{ }{
      \performTyp{
        \tarr{\zwsel{\htau_1}}{\htau_2}
      }{
        \aMove{\dParent}
      }{
        \zwsel{\tarr{\htau_1}{\htau_2}}
      }
    }
  \end{equation*}
  \gap{}
  \begin{equation*}
    \tag{\rname{TMArrParent2}}
    \inferrule{ }{
      \performTyp{
        \tarr{{\htau_1}}{\zwsel{\htau_2}}
      }{
        \aMove{\dParent}
      }{
        \zwsel{\tarr{\htau_1}{\htau_2}}
      }
    }
  \end{equation*}

  \paragraph{Type Deletion}
  \begin{equation*}
    \tag{\rname{TMDel}}
    \inferrule{ }{
      \performTyp{
        \zwsel{\htau}
      }{
        \aDel
      }{
        \zwsel{\tehole}
      }
    }
  \end{equation*}

  \paragraph{Type Construction}
  \begin{equation*}
    \tag{\rname{TMConArrow}}
    \inferrule{ }{
      \performTyp{
        \zwsel{\htau}
      }{
        \aConstruct{\farr}
      }{
        \tarr{\htau}{\zwsel{\tehole}}
      }
    }
  \end{equation*}
  \gap{}
  \begin{equation*}
    \tag{\rname{TMConNum}}
    \inferrule{ }{
      \performTyp{
        \zwsel{\tehole}
      }{
        \aConstruct{\fnum}
      }{
        \zwsel{\tnum}
      }
    }
  \end{equation*}

  \paragraph{Zipper Cases}
  \begin{equation*}
    \tag{\rname{TMArrZip1}}
    \inferrule{
      \performTyp{\ztau}{\alpha}{\ztau'}
    }{
      \performTyp{
        \tarr{\ztau}{\htau}
      }{
        \alpha
      }{
        \tarr{\ztau'}{\htau}
      }
    }
  \end{equation*}
  \gap{}
  \begin{equation*}
    \tag{\rname{TMArrZip2}}
    \inferrule{
      \performTyp{\ztau}{\alpha}{\ztau'}
    }{
      \performTyp{
        \tarr{\htau}{\ztau}
      }{
        \alpha
      }{
        \tarr{\htau}{\ztau'}
      }
    }
  \end{equation*}
\end{subequations}

\subsubsection{Expression Movement Actions}
\noindent\fbox{$\performMove{\zexp}{\aMove{\delta}}{\zexp'}$}

\begin{subequations}
  \paragraph{Ascription}
  \begin{equation*}
    \tag{\rname{EMAscChild1}}
    \inferrule{ }{
      \performTyp{
        \zwsel{\hexp : \htau}
      }{
        \aMove{\dChildn{1}}
      }{
        \zwsel{\hexp} : \htau
      }
    }
  \end{equation*}
  \gap{}
  \begin{equation*}
    \tag{\rname{EMAscChild2}}
    \inferrule{ }{
      \performTyp{
        \zwsel{\hexp : \htau}
      }{
        \aMove{\dChildn{2}}
      }{
        \hexp : \zwsel{\htau}
      }
    }
  \end{equation*}
  \gap{}
  \begin{equation*}
    \tag{\rname{EMAscParent1}}
    \inferrule{ }{
      \performTyp{
        \zwsel{\hexp} : \htau
      }{
        \aMove{\dParent}
      }{
        \zwsel{\hexp : \htau}
      }
    }
  \end{equation*}
  \gap{}
  \begin{equation*}
    \tag{\rname{EMAscParent2}}
    \inferrule{ }{
      \performTyp{
        \hexp : \zwsel{\htau}
      }{
        \aMove{\dParent}
      }{
        \zwsel{\hexp : \htau}
      }
    }
  \end{equation*}

  \paragraph{Lambda}
  \begin{equation*}
    \tag{\rname{EMLamChild1}}
    \inferrule{ }{
      \performMove{
        \zwsel{\hlam{x}{\hexp}}
      }{
        \aMove{\dChildn{1}}
      }{
        \hlam{x}{\zwsel{\hexp}}
      }
    }
  \end{equation*}
  \gap{}
  \begin{equation*}
    \tag{\rname{EMLamParent}}
    \inferrule{ }{
      \performMove{
        \hlam{x}{\zwsel{\hexp}}
      }{
        \aMove{\dParent}
      }{
        \zwsel{\hlam{x}{\hexp}}
      }
    }
  \end{equation*}

  \paragraph{Plus}
  \begin{equation*}
    \tag{\rname{EMPlusChild1}}
    \inferrule{ }{
      \performMove{
        \zwsel{\hadd{\hexp_1}{\hexp_2}}
      }{
        \aMove{\dChildn{1}}
      }{
        \hadd{\zwsel{\hexp_1}}{\hexp_2}
      }
    }
  \end{equation*}
  \gap{}
  \begin{equation*}
    \tag{\rname{EMPlusChild2}}
    \inferrule{ }{
      \performMove{
        \zwsel{\hadd{\hexp_1}{\hexp_2}}
      }{
        \aMove{\dChildn{2}}
      }{
        \hadd{\hexp_1}{\zwsel{\hexp_2}}
      }
    }
  \end{equation*}
  \gap{}
  \begin{equation*}
    \tag{\rname{EMPlusParent1}}
    \inferrule{ }{
      \performMove{
        \hadd{\zwsel{\hexp_1}}{\hexp_2}
      }{
        \aMove{\dParent}
      }{
        \zwsel{\hadd{\hexp_1}{\hexp_2}}
      }
    }
  \end{equation*}
  \gap{}
  \begin{equation*}
    \tag{\rname{EMPlusParent2}}
    \inferrule{ }{
      \performMove{
        \hadd{{\hexp_1}}{\zwsel{\hexp_2}}
      }{
        \aMove{\dParent}
      }{
        \zwsel{\hadd{\hexp_1}{\hexp_2}}
      }
    }
  \end{equation*}

  \paragraph{Application}
  \begin{equation*}
    \tag{\rname{EMApChild1}}
    \inferrule{ }{
      \performMove{
        \zwsel{\hap{\hexp_1}{\hexp_2}}
      }{
        \aMove{\dChildn{1}}
      }{
        \hap{\zwsel{\hexp_1}}{\hexp_2}
      }
    }
  \end{equation*}
  \gap{}
  \begin{equation*}
    \tag{\rname{EMApChild2}}
    \inferrule{ }{
      \performMove{
        \zwsel{\hap{\hexp_1}{\hexp_2}}
      }{
        \aMove{\dChildn{2}}
      }{
        \hap{\hexp_1}{\zwsel{\hexp_2}}
      }
    }
  \end{equation*}
  \gap{}
  \begin{equation*}
    \tag{\rname{EMApParent1}}
    \inferrule{ }{
      \performMove{
        \hap{\zwsel{\hexp_1}}{\hexp_2}
      }{
        \aMove{\dParent}
      }{
        \zwsel{\hap{\hexp_1}{\hexp_2}}
      }
    }
  \end{equation*}
  \gap{}
  \begin{equation*}
    \tag{\rname{EMApParent2}}
    \inferrule{ }{
      \performMove{
        \hap{{\hexp_1}}{\zwsel{\hexp_2}}
      }{
        \aMove{\dParent}
      }{
        \zwsel{\hap{\hexp_1}{\hexp_2}}
      }
    }
  \end{equation*}

  \paragraph{Non-Empty Hole}
  \begin{equation*}
    \tag{\rname{EMNEHoleChild1}}
    \inferrule{ }{
      \performMove{
        \zwsel{\hhole{\hexp}}
      }{
        \aMove{\dChildn{1}}
      }{
        \hhole{\zwsel{\hexp}}
      }
    }
  \end{equation*}
  \gap{}
  \begin{equation*}
    \tag{\rname{EMNEHoleParent}}
    \inferrule{ }{
      \performMove{
        \hhole{\zwsel{\hexp}}
      }{
        \aMove{\dParent}
      }{
        \zwsel{\hhole{\hexp}}
      }
    }
  \end{equation*}

\end{subequations}
\subsubsection{Synthetic and Analytic Expression Actions}
The synthetic and analytic expression action performance judgements are
defined mutually inductively.

\noindent\fbox{$\performSyn{\hGamma}{\zexp}{\htau}{\alpha}{\zexp'}{\htau'}$}

\begin{subequations}\label{Arules:performSyn}
  \paragraph{Movement}
  \begin{equation*}
    \tag{\rname{SAMove}}
    \inferrule{
      \performMove{\zexp}{\aMove{\delta}}{\zexp'}
    }{
      \performSyn{\hGamma}{\zexp}{\htau}{\aMove{\delta}}{\zexp'}{\htau}
    }
  \end{equation*}

  \paragraph{Deletion}
  \begin{equation*}
    \tag{\rname{SADel}}
    \inferrule{ }{
      \performSyn{\hGamma}{\zwsel{\hexp}}{\htau}{\aDel}{\zwsel{\hehole}}{\tehole}
    }
  \end{equation*}

  \paragraph{Construction}
  \begin{equation*}
    \tag{\rname{SAConAsc}}
    \inferrule{ }{
      \performSyn{\hGamma}{\zwsel{\hexp}}{\htau}{\aConstruct{\fasc}}{\hexp : \zwsel{\htau}}{\htau}
    }
  \end{equation*}
  \gap{}
  \begin{equation*}
    \tag{\rname{SAConVar}}
    \inferrule{ }{
      \performSyn{\hGamma, x : \htau}{\zwsel{\hehole}}{\tehole}{\aConstruct{\fvar{x}}}{\zwsel{x}}{\htau}
    }
  \end{equation*}
  \gap{}
  \begin{equation*}
    \tag{\rname{SAConLam}}
    \inferrule{ }{
      \performSyn
          {\hGamma}
          {\zwsel{\hehole}}
          {\tehole}
          {\aConstruct{\flam{x}}}
          {\hlam{x}{\hehole} : \tarr{\zwsel{\tehole}}{\tehole}}
          {\tarr{\tehole}{\tehole}}
    }
  \end{equation*}
  \gap{}
  \begin{equation*}
    \tag{\rname{SAConApArr}}
    \inferrule{
      \arrmatch{\htau}{\tarr{\htau_1}{\htau_2}}
    }{
      \performSyn
          {\hGamma}
          {\zwsel{\hexp}}
          {\htau}
          {\aConstruct{\fap}}
          {\hap{\hexp}{\zwsel{\hehole}}}
          {\htau_2}
    }
  \end{equation*}
  \gap{}
  \begin{equation*}
    \tag{\rname{SAConApOtw}}
    \inferrule{
      \tincompat{\htau}{\tarr{\tehole}{\tehole}}
    }{
      \performSyn
          {\hGamma}
          {\zwsel{\hexp}}
          {\htau}
          {\aConstruct{\fap}}
          {\hap{\hhole{\hexp}}{\zwsel{\hehole}}}
          {\tehole}
    }
  \end{equation*}
  \gap{}
  \begin{equation*}
    \tag{\rname{SAConNumLit}}
    \inferrule{ }{
      \performSyn
          {\hGamma}
          {\zwsel{\hehole}}
          {\tehole}
          {\aConstruct{\fnumlit{n}}}
          {\zwsel{\hnum{n}}}
          {\tnum}
    }
  \end{equation*}
  \gap{}
  \begin{equation*}
    \tag{\rname{SAConPlus1}}
    \inferrule{
      \tcompat{\htau}{\tnum}
    }{
      \performSyn
          {\hGamma}
          {\zwsel{\hexp}}
          {\htau}
          {\aConstruct{\fplus}}
          {\hadd{\hexp}{\zwsel{\hehole}}}
          {\tnum}
    }
  \end{equation*}
  \gap{}
  \begin{equation*}
    \tag{\rname{SAConPlus2}}
    \inferrule{
      \tincompat{\htau}{\tnum}
    }{
      \performSyn
          {\hGamma}
          {\zwsel{\hexp}}
          {\htau}
          {\aConstruct{\fplus}}
          {\hadd{\hhole{\hexp}}{\zwsel{\hehole}}}
          {\tnum}
    }
  \end{equation*}
  \gap{}
  \begin{equation*}
    \tag{\rname{SAConNEHole}}
    \inferrule{ }{
      \performSyn
          {\hGamma}
          {\zwsel{\hexp}}
          {\htau}
          {\aConstruct{\fnehole}}
          {\hhole{\zwsel{\hexp}}}
          {\tehole}
    }
  \end{equation*}

  \paragraph{Finishing}
  \begin{equation*}
    \tag{\rname{SAFinish}}
    \inferrule{
      \hsyn{\hGamma}{\hexp}{\htau'}
    }{
      \performSyn
          {\hGamma}
          {\zwsel{\hhole{\hexp}}}
          {\tehole}
          {\aFinish}
          {\zwsel{\hexp}}
          {\htau'}
    }
  \end{equation*}

  \paragraph{Zipper Cases}
  \begin{equation*}
    \tag{\rname{SAZipAsc1}}
    \inferrule{
      \performAna
          {\hGamma}
          {\zexp}
          {\htau}
          {\alpha}
          {\zexp'}
    }{
      \performSyn
          {\hGamma}
          {\zexp : \htau}
          {\htau}
          {\alpha}
          {\zexp' : \htau}
          {\htau}
    }
  \end{equation*}
  \gap{}
  \begin{equation*}
    \tag{\rname{SAZipAsc2}}
    \inferrule{
      \performTyp{\ztau}{\alpha}{\ztau'}\\
      \hana{\hGamma}{\hexp}{\removeSel{\ztau'}}
    }{
      \performSyn
          {\hGamma}
          {\hexp : \ztau}
          {\removeSel{\ztau}}
          {\alpha}
          {\hexp : \ztau'}
          {\removeSel{\ztau'}}
    }
  \end{equation*}
  \gap{}
  \begin{equation*}
    \tag{\rname{SAZipApArr}}
    \inferrule{
      \hsyn{\hGamma}{\removeSel{\zexp}}{\htau_2}\\
      \performSyn
          {\hGamma}
          {\zexp}
          {\htau_2}
          {\alpha}
          {\zexp'}
          {\htau_3}\\\\
          \arrmatch{\htau_3}{\tarr{\htau_4}{\htau_5}}\\
          \hana{\hGamma}{\hexp}{\htau_4}
    }{
      \performSyn
          {\hGamma}
          {\hap{\zexp}{\hexp}}
          {\htau_1}
          {\alpha}
          {\hap{\zexp'}{\hexp}}
          {\htau_5}
    }
  \end{equation*}
  \gap{}
  \begin{equation*}
    \tag{\rname{SAZipApAna}}
    \inferrule{
      \hsyn{\hGamma}{\hexp}{\htau_2}\\
      \arrmatch{\htau_2}{\tarr{\htau_3}{\htau_4}}\\
      \performAna
          {\hGamma}
          {\zexp}
          {\htau_3}
          {\alpha}
          {\zexp'}
    }{
      \performSyn
          {\hGamma}
          {\hap{\hexp}{\zexp}}
          {\htau_1}
          {\alpha}
          {\hap{\hexp}{\zexp'}}
          {\htau_4}
    }
  \end{equation*}
  \gap{}
  \begin{equation*}
    \tag{\rname{SAZipPlus1}}
    \inferrule{
      \performAna
          {\hGamma}
          {\zexp}
          {\tnum}
          {\alpha}
          {\zexp'}
    }{
      \performSyn
          {\hGamma}
          {\hadd{\zexp}{\hexp}}
          {\tnum}
          {\alpha}
          {\hadd{\zexp'}{\hexp}}
          {\tnum}
    }
  \end{equation*}
  \gap{}
  \begin{equation*}
    \tag{\rname{SAZipPlus2}}
    \inferrule{
      \performAna
          {\hGamma}
          {\zexp}
          {\tnum}
          {\alpha}
          {\zexp'}
    }{
      \performSyn
          {\hGamma}
          {\hadd{\hexp}{\zexp}}
          {\tnum}
          {\alpha}
          {\hadd{\hexp}{\zexp'}}
          {\tnum}
    }
  \end{equation*}
  \gap{}
  \begin{equation*}
    \tag{\rname{SAZipHole}}
    \inferrule{
      \hsyn{\hGamma}{\removeSel{\zexp}}{\htau}\\
      \performSyn
          {\hGamma}
          {\zexp}
          {\htau}
          {\alpha}
          {\zexp'}
          {\htau'}
    }{
      \performSyn
          {\hGamma}
          {\hhole{\zexp}}
          {\tehole}
          {\alpha}
          {\hhole{\zexp'}}
          {\tehole}
    }
  \end{equation*}
\end{subequations}

\noindent\fbox{$\performAna{\hGamma}{\zexp}{\htau}{\alpha}{\zexp'}$}
\begin{subequations}\label{Arules:performAna}
  \paragraph{Subsumption}
  \begin{equation*}
    \tag{\rname{AASubsume}}
    \inferrule{
      \hsyn{\hGamma}{\removeSel{\zexp}}{\htau'}\\
      \performSyn{\hGamma}{\zexp}{\htau'}{\alpha}{\zexp'}{\htau''}\\
      \tcompat{\htau}{\htau''}
    }{
      \performAna{\hGamma}{\zexp}{\htau}{\alpha}{\zexp'}
    }
  \end{equation*}

  \paragraph{Movement}
  \begin{equation*}
    \tag{\rname{AAMove}}
    \inferrule{
      \performMove{\zexp}{\aMove{\delta}}{\zexp'}
    }{
      \performAna{\hGamma}{\zexp}{\htau}{\aMove{\delta}}{\zexp'}
    }
  \end{equation*}

  \paragraph{Deletion}
  \begin{equation*}
    \tag{\rname{AADel}}
    \inferrule{ }{
      \performAna{\hGamma}{\zwsel{\hexp}}{\htau}{\aDel}{\zwsel{\hehole}}
    }
  \end{equation*}

  \paragraph{Construction}
  \begin{equation*}
    \tag{\rname{AAConAsc}}
    \inferrule{ }{
      \performAna{\hGamma}{\zwsel{\hexp}}{\htau}{\aConstruct{\fasc}}{\hexp : \zwsel{\htau}}
    }
  \end{equation*}
  \gap{}
  \begin{equation*}
    \tag{\rname{AAConVar}}
    \inferrule{
      \tincompat{\htau}{\htau'}
    }{
      \performAna{\hGamma, x : \htau'}{\zwsel{\hehole}}{\htau}{\aConstruct{\fvar{x}}}{\hhole{\zwsel{x}}}
    }
  \end{equation*}
  \gap{}
  \begin{equation*}
    \tag{\rname{AAConLam1}}
    \inferrule{
      \arrmatch{\htau}{\tarr{\htau_1}{\htau_2}}
    }{
      \performAna
          {\hGamma}
          {\zwsel{\hehole}}
          {\htau}
          {\aConstruct{\flam{x}}}
          {\hlam{x}{\zwsel{\hehole}}}
    }
  \end{equation*}
  \gap{}
  \begin{equation*}
    \tag{\rname{AAConLam2}}
    \inferrule{
      \tincompat{\htau}{\tarr{\tehole}{\tehole}}
    }{
      \performAna
          {\hGamma}
          {\zwsel{\hehole}}
          {\htau}
          {\aConstruct{\flam{x}}}
          {\hhole{
              \hlam{x}{\hehole} : \tarr{\zwsel{\tehole}}{\tehole}
          }}
    }
  \end{equation*}
  \gap{}
  \begin{equation*}
    \tag{\rname{AAConNumLit}}
    \inferrule{
      \tincompat{\htau}{\tnum}
    }{
      \performAna
          {\hGamma}
          {\zwsel{\hehole}}
          {\htau}
          {\aConstruct{\fnumlit{n}}}
          {\hhole{\zwsel{\hnum{n}}}}
    }
  \end{equation*}

  \paragraph{Finishing}
  \begin{equation*}
    \tag{\rname{AAFinish}}
    \inferrule{
      \hana{\hGamma}{\hexp}{\htau}
    }{
      \performAna
          {\hGamma}
          {\zwsel{\hhole{\hexp}}}
          {\htau}
          {\aFinish}
          {\zwsel{\hexp}}
    }
  \end{equation*}

  \paragraph{Zipper Cases}
  \begin{equation*}
    \tag{\rname{AAZipLam}}
    \inferrule{
      \arrmatch{\htau}{\tarr{\htau_1}{\htau_2}}\\
      \performAna
          {\hGamma, x : \htau_1}
          {\zexp}
          {\htau_2}
          {\alpha}
          {\zexp'}
    }{
      \performAna
          {\hGamma}
          {\hlam{x}{\zexp}}
          {\htau}
          {\alpha}
          {\hlam{x}{\zexp'}}
    }
  \end{equation*}
\end{subequations}
\subsubsection{Iterated Action Judgements} ~

\noindent $\mathsf{ActionList}$~~$\bar{\alpha} ::= \cdot ~\vert~ \alpha; \bar{\alpha}$\vspace{4px}\\
\fbox{$\performTypI{\ztau}{\bar{\alpha}}{\ztau'}$}
\begin{subequations}
  \begin{equation*}
    \tag{\rname{DoRefl}}
    \inferrule{ }{
      \performTypI{\ztau}{\cdot}{\ztau}
    }
  \end{equation*}
  \gap{}
  \begin{equation*}
    \tag{\rname{DoType}}
    \inferrule{
      \performTyp{\ztau}{\alpha}{\ztau'}\\
      \performTypI{\ztau'}{\bar{\alpha}}{\ztau''}
    }{
      \performTypI{\ztau}{\alpha; \bar{\alpha}}{\ztau''}
    }
  \end{equation*}
\end{subequations}

\noindent \fbox{$\performSynI{\hGamma}{\zexp}{\htau}{\bar{\alpha}}{\zexp'}{\htau'}$}
\begin{subequations}
  \begin{equation*}
    \tag{\rname{DoRefl}}
    \inferrule{ }{
      \performSynI{\hGamma}{\zexp}{\htau}{\cdot}{\zexp}{\htau}
    }
  \end{equation*}
  \gap{}
  \begin{equation*}
    \tag{\rname{DoSynth}}
    \inferrule{
      \performSyn{\hGamma}{\zexp}{\htau}{\alpha}{\zexp'}{\htau'}\\
      \performSynI{\hGamma}{\zexp'}{\htau'}{\bar{\alpha}}{\zexp''}{\htau''}
    }{
      \performSynI{\hGamma}{\zexp}{\htau}{\alpha; \bar{\alpha}}{\zexp''}{\htau''}
    }
  \end{equation*}
\end{subequations}

\noindent \fbox{$\performAnaI{\hGamma}{\zexp}{\htau}{\bar{\alpha}}{\zexp'}$}
\begin{subequations}
  \begin{equation*}
    \tag{\rname{DoRefl}}
    \inferrule{ }{
      \performAnaI{\hGamma}{\zexp}{\htau}{\cdot}{\zexp}
    }
  \end{equation*}
  \gap{}
  \begin{equation*}
    \tag{\rname{DoAna}}
    \inferrule{
      \performAna{\hGamma}{\zexp}{\htau}{\alpha}{\zexp'}\\
      \performAnaI{\hGamma}{\zexp'}{\htau}{\bar\alpha}{\zexp''}
    }{
      \performAnaI{\hGamma}{\zexp}{\htau}{\alpha; \bar\alpha}{\zexp''}
    }
  \end{equation*}
\end{subequations}

\noindent \fbox{$\bar\alpha~\mathsf{movements}$}
\begin{subequations}
  \begin{equation*}
    \tag{\rname{AM[]}}
    \inferrule{ }{
      \cdot~\mathsf{movements}
    }
  \end{equation*}
  \gap{}
  \begin{equation*}
    \tag{\rname{AM::}}
    \inferrule{
      \bar\alpha~\mathsf{movements}
    }{
      \aMove{\delta}; \bar\alpha~\mathsf{movements}
    }
  \end{equation*}
\end{subequations}

\else
No Appendix Here
\fi

\end{document}